\documentclass[pra,twocolumn,aps,superscriptaddress,showpacs,longbibliography]{revtex4-1}
\usepackage{hyperref}
\hypersetup{
	colorlinks=true,
	citecolor=blue,
	linkcolor=blue,
	urlcolor=blue}
\usepackage{graphicx}
\usepackage{amsmath}
\usepackage{physics}
\usepackage{amsfonts}
\usepackage{amssymb}
\usepackage{bm}
\usepackage{siunitx}
\usepackage{xfrac}
\usepackage[normalem]{ulem}
\usepackage{braket}% Dirac braket
\usepackage{cases}% Numbered cases
\newcommand{\changes}[1]{\textcolor{black}{ #1}}

\begin{document}
%\linenumbers
%%%%%%%%%%%%%%%%%%%%%%%%%%%%%%%%%%%%%%%%%%%%%%%%%%%%%%%%%%%%%%%%%%%%%%%%%%%%%%%
\title[]{Finite temperature ferromagnetic transition in coherently coupled Bose gases}

\author{Arko Roy}
\address{Pitaevskii BEC Center, CNR-INO and Dipartimento di Fisica, Universit\`a di Trento, via Sommarive 14, I-38123 Trento, Italy} 
\address{School of Physical Sciences, Indian Institute of Technology Mandi, Mandi-175075 (H.P.), India}

\author{Miki Ota}
\address{Pitaevskii BEC Center, CNR-INO and Dipartimento di Fisica, Universit\`a di Trento, via Sommarive 14, I-38123 Trento, Italy} 

\author{Franco Dalfovo}
\address{Pitaevskii BEC Center, CNR-INO and Dipartimento di Fisica, Universit\`a di Trento, via Sommarive 14, I-38123 Trento, Italy}

\author{Alessio Recati}
\address{Pitaevskii BEC Center, CNR-INO and Dipartimento di Fisica, Universit\`a di Trento, via Sommarive 14, I-38123 Trento, Italy} 
\address{
Trento Institute for Fundamental Physics and Applications, INFN, 38123 Povo, Italy}

%\pagewiselinenumbers

%%%%%%%%%%%%%%%%%%%%%%%%%%%%%%%%%%%%%%%%%%%%%%%%%%%%%%%%%%%%%%%%%%%%%%%%%%%%%
%%%%%%%%%%%%%                  Abstract                        %%%%%%%%%%%%%%
%%%%%%%%%%%%%%%%%%%%%%%%%%%%%%%%%%%%%%%%%%%%%%%%%%%%%%%%%%%%%%%%%%%%%%%%%%%%%

\begin{abstract}
A paramagnetic-ferromagnetic quantum phase transition is known to occur at zero temperature in a two-dimensional coherently-coupled Bose mixture of dilute ultracold atomic gases provided the interspecies interaction strength is large enough. Here we study the fate of such a transition at finite temperature by performing numerical simulations with the stochastic (projected) Gross-Pitaevskii formalism, which includes both thermal and beyond mean-field effects. By extracting the average magnetization, the magnetic fluctuations and characteristic relaxation frequency (or, critical slowing down), we identify a finite temperature critical line for the transition. We find that the critical point shifts linearly with  temperature and, in addition, the three quantities used to probe the transition exhibit a temperature power-law scaling. The scaling of the critical slowing down is found to be consistent with thermal critical exponents and is very well approximated by the square of the spin excitation gap at the zero-temperature.
\end{abstract}

\maketitle

%%%%%%%%%%%%%%%%%%%%%%%%%%%%%%%%%%%%%%%%%%%%%%%%%%%%%%%%%%%%%%%
%              Section: Introduction                 
%%%%%%%%%%%%%%%%%%%%%%%%%%%%%%%%%%%%%%%%%%%%%%%%%%%%%%%%%%%%%%%

\section{Introduction}
\label{Introduction}

One of the most quintessential example of quantum-phase transition in cold atomic systems is the magnetic transition in spinor gases~\cite{stamperkurn_13}. Spinor gases are multi-component systems of degenerate quantum gases, where the spin degree of freedom arises from the internal spin of the constituent atoms~\cite{pethick_08,kawaguchi,pitaevskii-16}. These systems have been the object of profound theoretical~\cite{ho_96,timmermans_98,ao_98,trippenbach_2000,schaeybroeck_08,wen_12} and experimental~\cite{modugno_02,thalhammer_08,lercher_11,mccarron_11,pasquiou_13,papp_08,tojo_10,wilson_21,warner_21} investigation over the last two decades. In particular, the possibility to experimentally realize quantum mixtures by utilizing the same atomic species in different internal states opens up a plethora of striking opportunities to perform quantum emulation of magnetic materials~\cite{farolfi-2020}, and is indeed reminiscent of the paradigmatic Ising model in condensed matter physics~\cite{sachdev_2011,zurek_05}. The components here share the same statistics, and the change in the initial populations  can occur, for instance, through spin-changing collisions or via coherent coupling~\cite{abad_13}. The spin degree of freedom allows for the realization of multi-component vector order parameter, with characteristics typical of both superfluid and magnetic systems, such as quantum phase coherence, long-range order and symmetry breaking through the presence of different zero-temperature phases~\cite{kawaguchi_12,stamperkurn_13}.
With alkali and alkali-earth atoms, most mixtures of sub-states of the same hyperfine manifold are long-lived and allow for the study of spinor gases with pseudo-spin-1/2~\cite{hall_98,myatt_97,barrett_01,semeghini_18}, spin-1~\cite{stenger_98,kang_19,bookjans_11,luo_17} and spin-2~\cite{chang_04,schmaljohann_04} configuration. These states can be coherently manipulated via optical or radiofrequency fields~\cite{zhai_12,zhai_2015,lin_11}, making them suitable candidates to explore the role of symmetry and topology in
quantum materials~\cite{kawaguchi_10}, quantum phase transitions~\cite{zibold_10,nicklas_15},
non-equilibrium quantum dynamics~\cite{kasamatsu_05,nicklas_11,zibold_10}, and the entanglement and squeezing of quantum fields~\cite{gross_11,lucke_11,takumi_21}, and analogues of quantum gravity models~\cite{fischer_04,garay_2000}, to mention a few.

For any quantum phase-transition, the most dramatic manifestation of critical phenomena is the appearance of divergent fluctuations of collective observables at the point of phase-transition. Although occurring at zero temperature, the quantum critical phenomena are also sensitive to thermal effects, which can be probed in temperature ranges accessible in experiments~\cite{hazzard_11, christensen_21}. Non-zero temperature enlarges the parameter space of the quantum critical point, the broadened region constituting the quantum critical region~\cite{sondhi_97,sachdev_2011,dutta_2015,carr_10}.
Relevant for the present work, the critical point of phase-transition in a system of coherently coupled ultracold bosons with total density $n=n_1+n_2$, is entirely determined by the $s$-wave scattering length, $a$, and the strength of the Rabi coupling $\Omega$. In particular, in an interacting mixture of two-components at zero-temperature, the para-ferromagnetic transition occurs at $g_{12} = g + 2\Omega/n$ with $g_{12}$ as the intercomponent, and $g$ as the intracomponent coupling constants~\cite{abad_13}. At finite temperature, deviations from this equality are expected to occur, and while the spin-orbit coupled Bose gases have gathered both theoretical~\cite{chen_17,chen_18,su_17,attanasio_20,liu_12,su_12} and experimental~\cite{ji_14} interest, the coherently-coupled condensates have received less attention, despite being one of the simplest, yet rich, implementations of a spinor condensate with an external field. The current work aims to fill this gap, by investigating the finite-temperature effects on the ferromagnetic phase-transition.

To proceed with our investigation, we consider a two-dimensional homogeneous system of an atomic species occupying two different hyperfine states which are coherently coupled by an external radiation field. This composite system generalizes the well-known idea of Rabi oscillation in quantum optics~\cite{gerry_knight_2004} to extended non-linear systems, and can also be described in terms of internal Josephson dynamics~\cite{leggett_01,farolfi_21}. With the advancement in experimental techniques, quasi-uniform box traps are available~\cite{gaunt_13,chomaz_15}, and thus it is timely to explore the phase-transition region in a system of coherently coupled condensates. Compared to a three-dimensional system, in which most of the time only column density can be measured, the feasibility of a two-dimensional (2D) planar configuration to probe local density fluctuations is much better.

To accomplish the study of the different phases of coherently coupled condensates, at the outset, we compute the hysteresis curves for the admissible phases by invoking the detuning parameter in the mean-field Gross-Pitaevskii formalism. We then move our attention to discuss the physics around the critical point using the Stochastic (projected) Gross-Pitaevskii model. We calculate the equilibrium magnetization, magnetization fluctuation, and demonstrate slowing of equilibration time of the composite system at different temperatures. The simulations indeed show evidences of the enhanced fluctuations at the critical point, and a deviation of the latter from its zero-temperature counterpart. 

\section{Homogeneous coherently coupled condensates at zero temperature}

\subsection{Formalism}

We consider a dilute, homogeneous, weakly interacting atomic Bose mixture in 2D at zero temperature whose atoms can occupy two different
hyperfine states $|\!\uparrow\rangle$ and $|\!\downarrow\rangle$, hereafter called as $1$ and $2$, separated by an energy $h\nu$. The atoms, with mass $m$, interact with each other via $s$-wave scattering with intra- and inter-species interaction strengths as $g_{11}$, $g_{22}$ and $g_{12}$, respectively. In addition, the two states can be coupled by means of an external coherent drive inducing a Rabi transfer of atoms between the two internal levels. This can be experimentally realized, for instance, by using a two-photon transition~\cite{farolfi_21}, characterized by the strength $\Omega$, which represents the intensity of the coupling of the atoms with the external electromagnetic field (here taken to be real and positive), and the detuning $\delta$, which is the difference between the frequency splitting $\nu$  and the frequency associated to the two-photon coupling.
Within the mean-field formalism, the zero temperature order parameters of the two components, $\psi_1({\bf x},t)$ and $\psi_2({\bf x},t)$ with ${\bf x}=(x,y)$, obey the following coupled Gross-Pitaevskii equations~\cite{pitaevskii_16,bernier_14}  
\begin{eqnarray}
 i\hbar\frac{\partial}{\partial t}\psi_i &=& \bigg[-\frac{\hbar^2 \nabla^2}{2m}
        + g_{ii}|\psi_{i}|^2  
        + g_{12}|\psi_{3-i}|^2 \nonumber\\ 
        &+& (-1)^i\delta\bigg]\psi_i +\Omega\psi_{3-i} \ ,
\label{cc1}
\end{eqnarray}
where $i={1,2}$. The number of atoms in each state is given by  $N_i= \int n_i({\bf x})\,d{\bf x}$, where $n_i=|\psi_i|^2$ are the corresponding densities. Due to the presence of the coherent Rabi coupling 
term, only the total number of atoms $N = N_1 + N_2$ is conserved giving rise to a $U(1)$ symmetry, unlike in uncoupled Bose-Bose mixtures ($\Omega=0$), where the number of particles in each individual component ($N_1$ and $N_2$) is conserved. We define $n=N/{\mathcal A}$ as the total number density with ${\mathcal A}$ as the area of the 2D box. Furthermore, we consider a symmetric interaction potential with $g_{11}=g_{22}=g$ and zero detuning giving rise to an additional $\mathbb{Z}_2$ symmetry corresponding to the exchange of components~\cite{abad_13}.

\begin{figure}[!hbtp]
\centering
\includegraphics[width=0.95\columnwidth]{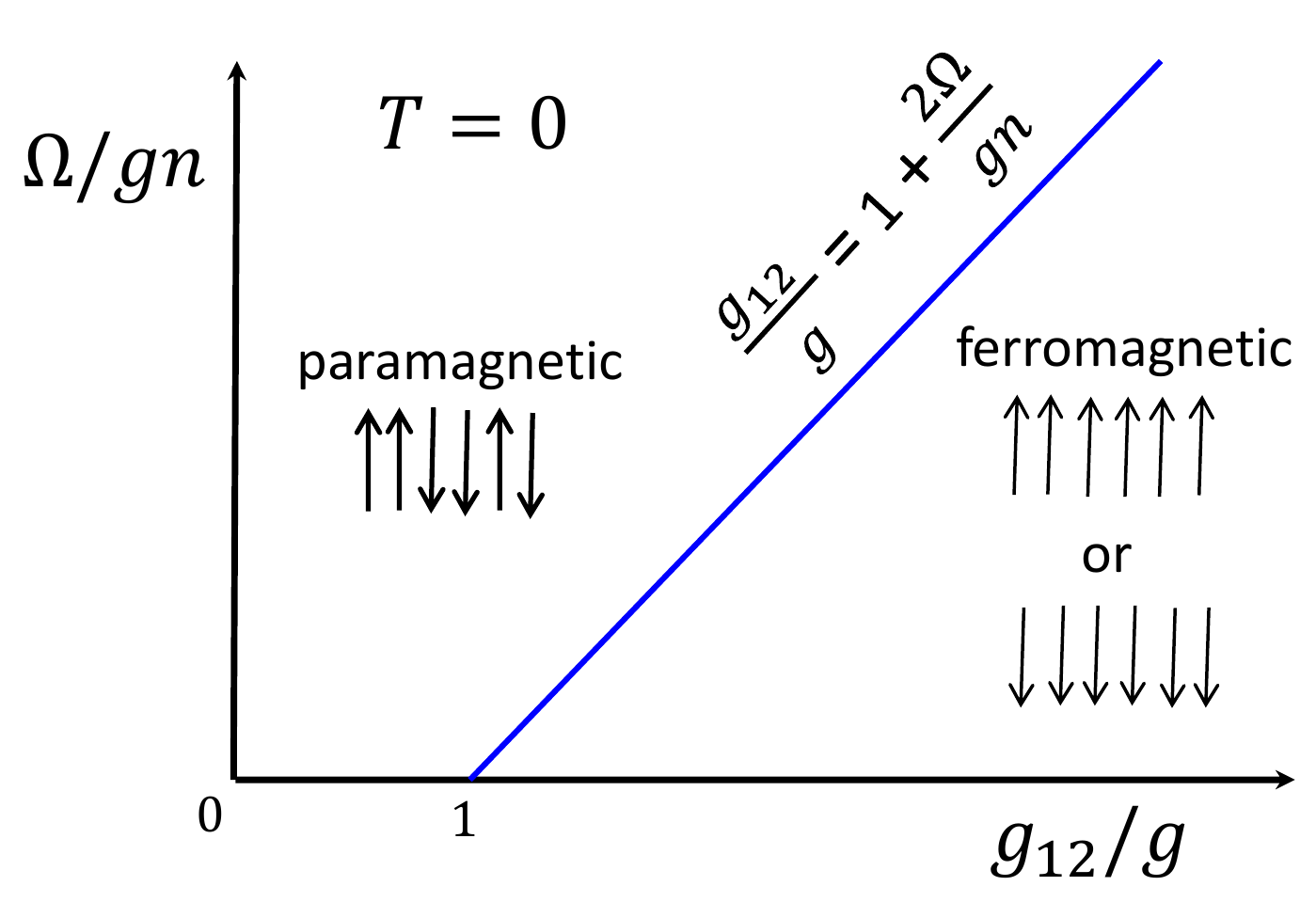}
\caption{Schematic of the phase diagram of the ground state of a homogeneous coherently coupled condensate $(\Omega\neq 0)$ at zero temperature. The solid blue line separates the paramagnetic from the ferromagnetic regime. For $T\neq 0$ and $\Omega\neq 0$, the transition is expected to get shifted and broadened. A detailed analysis for uncoupled ($\Omega=0$) Bose condensed mixtures at $T\neq0$ has been carried out in our previous work~\cite{roy_21}. }
\label{pf}
\end{figure}

\subsection{Characterization of phases}

The ground state of a coherently coupled homogeneous gas has uniform densities $n_1$ and $n_2$ and corresponds to the stationary solution of the coupled Gross-Pitaevskii equations (\ref{cc1}) with lowest energy. In the absence of detuning $(\delta=0)$, it corresponds to the minimum of the mean-field energy density~\cite{abad_13}
\begin{equation}
    \epsilon = \frac{1}{4}(g+g_{12})n^2 + \frac{1}{4}(g-g_{12})s_z^2 - \Omega\sqrt{n^2-s_z^2}-\mu n \ ,
    \label{energy}
\end{equation}
where $n=n_1+n_2$ and $s_z = n_1-n_2$ are the total density and the spin density, respectively. The chemical potential $\mu$, which is the same for both components, can be obtained by minimizing the above energy density with respect to the density $n$. One finds 
\begin{equation}
    \mu = \frac{n}{2}\bigg( g + g_{12} - \frac{\Omega}{\sqrt{n_1n_2}} \bigg) \, .
\end{equation}
Instead, by minimizing Eq.~(\ref{energy}) with respect to the spin density $s_z$, one obtains the equation
\begin{equation}
    s_z\bigg(g-g_{12} + \frac{2\Omega}{\sqrt{n^2-s_z^2}}\bigg) = 0 \ .
    \label{mfmag}
\end{equation}
The admissible solutions of this equation are governed by the values of the interaction strengths. In particular, the ground state of the system can either be a neutral paramagnetic phase, with $s_z=0$ and $U(1) \times \mathbb{Z}_2$ symmetry, or a spin polarized ferromagnetic phase, with $s_z = \pm n\sqrt{1-4\Omega^2/[(g-g_{12})n]^2}$ and broken $\mathbb{Z}_2$ symmetry. Defining the parameter ${\bar g}_{12} = g + 2\Omega/n$, the para-ferromagnetic phase transition is identified when $g_{12} = {\bar g}_{12}$. For  $g_{12} < {\bar g}_{12}$ the system is paramagnetic, while for $g_{12} > {\bar g}_{12}$ it is ferromagnetic. For a complete review of coherently coupled mixtures of condensates of dilute atomic gases we refer the reader to Refs.~\cite{abad_13,recati_19,recati_21}.
A schematic of the $T=0$ phase-diagram is shown in Fig.~\ref{pf}.

\begin{figure}[!hbtp]
\centering
\includegraphics[width=0.95\columnwidth]{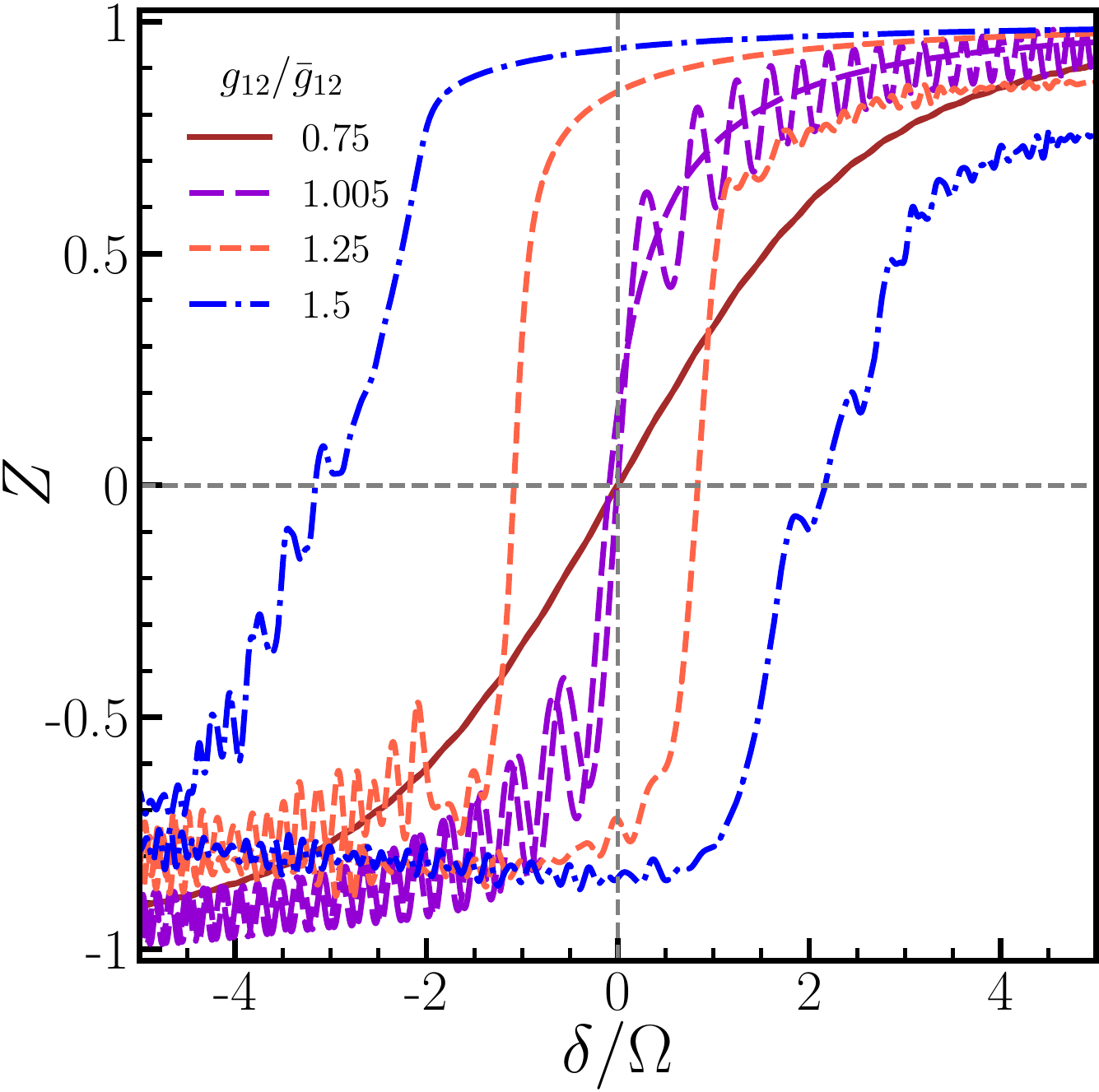}
\caption{Magnetization $Z = (N_1-N_2)/(N_1+N_2)$ vs. relative detuning $\delta/\Omega$, calculated by solving the Gross-Pitaevskii equation (\ref{cc1}) at $T=0$ for a cycle in which the value of $\delta$ is changed in time from $\delta_0$ to $-\delta_0$ and then back to $\delta_0$ again, at the same rate. Here we fix $\delta_0/\Omega=5$ and $\Omega=0.1gn$. If the initial configuration is paramagnetic, as for $g_{12}/{\bar g}_{12}=0.75$, the magnetization curve is the same in both directions and shows no magnetization at zero detuning. Conversely, if the initial configuration is ferromagnetic, as for $g_{12}/{\bar g}_{12}>1$, it exhibits an hysteresis curve with finite magnetization at zero detuning.  }
\label{hyst}
\end{figure}

The breaking of the discrete $\mathbb{Z}_2$ symmetry in the paramagnetic to ferromagnetic transition can be validated on solving Eq.~(\ref{cc1}) by invoking the detuning parameter $\delta$ and looking at the hysteresis loop exhibited by the ferromagnetic phase on being driven from positive to negative detuning values. To demonstrate this, we prepare the system with an initial detuning $\delta_0/\Omega = 5$ and with different values of $g_{12}/{\bar g}_{12}$ at $T=0$.  Then we vary $\delta$ from $\delta_0$ to $-\delta_0$ over time, and then again back to $\delta_0$, and we calculate the magnetization $Z=(N_1-N_2)/(N_1+N_2)$ during the cycle. If we start from a paramagnetic ground state ($g_{12}/{\bar g}_{12}<1$), the magnetization curve follows the same trend as the change in $\delta$ and retraces over the same path, without hysteresis, as shown in Fig.~\ref{hyst}. Instead, if we start from a ferromagnetic ground state ($g_{12}/{\bar g}_{12}>1$), the magnetization forms a hysteresis loop retaining a finite magnetization, $Z \neq 0$, even when $\delta=0$. The area of the loop increases with increasing $g_{12}/{\bar g}_{12}$.
We note on passing that similar hysteresis loops with cold atoms have been experimentally observed in a double-well potential setup exhibiting bifurcation~\cite{trenkwalder_16}. 
Very recently, the characterisation of hysteresis loops revealing the
para-ferromagnetic transition in a cigar-shaped coherently coupled condensate of sodium atoms has been reported in~\cite{cominotti_22}. 

After having characterized the properties of the coherently coupled gases away from the $\mathbb{Z}_2$ symmetry breaking point, we now investigate the region around it. In such a quantum critical region, characterized by enhanced fluctuations, any mean-field approach is expected to fail.

\begin{figure*}[!hbtp]
\centering
\includegraphics[width=\textwidth]{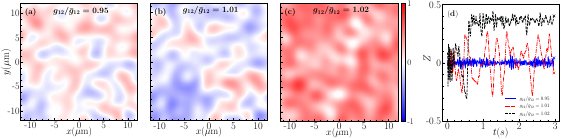}
\caption{\changes{Panels (a),(b),(c) show typical magnetization density profiles obtained by evolving Eq.~(\ref{sgpe}) for $3$~s, with zero detuning ($\delta=0$). Each simulation starts from a purely random $c$-field. Temperature is the same, $T/T_g=0.2$, while the interaction parameter is  $g_{12}/{\bar g}_{12}=0.95, 1.01$ and $1.02$, respectively. The color scale refers to the quantity $(n_1-n_2)/(n_1+n_2)$. Panel (d) shows the magnetization $Z$ during the full time evolution. The markers at the end of the trajectories are in correspondence of the snapshots in (a)-(c).}}
\label{oscnum}
\end{figure*}

\section{Stochastic Gross-Pitaevskii Formalism for coupled condensates}\label{model}
 
With having the need of a reliable theoretical description to investigate the physics around the critical point, we resort to the
Stochastic (projected) Gross-Pitaevskii formalism (SGPE)~\cite{stoof_01,proukakis_06,proukakis_08,blakie_08,bradley_08,cockburn_09,su_11,rooney_13,davis_13,berloff_14,brewczyk_2007,gallucci_16,kobayashi_16,ota_18}
adapted for multicomponent condensates~\cite{bradley_14,su_17,liu_12,su_12}. This framework describes the system and its fluctuations by using  a single noisy classical field coupled to a thermal bath, and also includes physical effects that are beyond mean-field theory. In the presence of detuning, the equations are given by
\begin{eqnarray}
i\hbar\frac{\partial}{\partial t}\psi_i({\bf x},t)&=& {\mathcal {\hat P}}\bigg\{
        (1-i\gamma)\bigg[\bigg(-\frac{\hbar^2 \nabla^2}{2m}
        + g|\psi_{i}({\bf x},t)|^2  \nonumber\\
        &+& g_{12}|\psi_{3-i}|^2 +(-1)^i\delta - \mu\bigg)\psi_i+ \Omega\psi_{3-i}\bigg] \nonumber\\
        &+& \eta_i({\bf x},t)\bigg\} \ .
\label{sgpe}
\end{eqnarray}
The two complex functions $\psi_{i}({\bf x},t)$, representing the ``classical" fields ($c$-fields), account for the macroscopically occupied low-energy modes of each component of the gas (labelled by the index $i \in \{1,2\}$) subject to random thermal fluctuations.  The corresponding densities are $n_i({\bf x},t)=|\psi_i({\bf x},t)|^2$. The $c$-fields $\psi_{i}({\bf x},t)$ include the multi-mode coherent region of the energy spectrum up to an energy cutoff $\epsilon_{{\rm cut}}$. The energy cut-off is chosen as~\cite{blakie_08,proukakis_08,rooney_10,comaron_19,fabrizio_18,liu_20}
\begin{equation}
	\epsilon_{{\rm cut}} = k_{\rm B} T\ln 2 + \mu \, .
	\label{ecut}
\end{equation}
where $\mu$ is the chemical potential. This choice guarantees that the mean occupation of the modes below $\epsilon_{{\rm cut}}$ is larger than unity, but the precise value of the cutoff is not crucial, as long as it belongs to a reasonable range. The same choice of the cutoff has been earlier used to validate experimental results for single component ~\cite{ota_18,comaron_19,fabrizio_18} and two-component~\cite{roy_21} condensates in similar configurations. The projector $\mathcal {\hat P}$ compels the $c$-fields to lie within the coherent region at each time-step. 

The modes above the cut-off represent the incoherent region of the energy spectrum; it is the source of a stochastic Gaussian random noise which satisfies the following fluctuation-dissipation theorem
\begin{equation}\label{Eq:Gaussian-noise}
	\langle \eta_i({\bf x},t) \eta_j^*({\bf x}',t')\rangle = 2\hbar \gamma k_{\rm B} T  \delta({\bf x} - {\bf x}')\delta(t-t')\delta_{ij} \, ,
\end{equation}
where $\langle \cdots \rangle$ denotes the averaging over different noise realizations. Following Refs.~\cite{su_17,liu_12,su_12}, related to spin-orbit coupled Bose gases, the noise terms in the present study are taken to be independent for the coupled $c$-field densities. 
The amount of coupling between the coherent and incoherent regions is fixed by the parameter $\gamma$, which accounts for the thermal equilibration rate. In this work, we choose $\gamma= 0.01$, which is the same of Ref.~\cite{ota_18}. Similar values were also used in  \cite{comaron_19} and \cite{liu2018}; in the latter case, the parameter $\gamma$ was optimized to reproduce typical experimental growth rates of single component condensates in 3D. It is to be noted that in SGPE individual results obtained with independent noise realizations \changes{can be (arguably, see, e.g.~\cite{sakmann_16,sakmann_17,olsen-17a} and references therein) thought of as being equivalent to the individual results obtained from independent experimental runs;} due to the random nature of the noise, the outcomes of each noise realizations  will differ  from  one  another  as  is  the  case in experiments. 

\changes{In order to perform simulations which are sensible for feasible experiments~\cite{gaunt_13,ville_18,farolfi_21}, we confine the gas in a box potential in the $x$-$y$ plane and harmonic trap in the $z$-direction. We use a uniform 2D square box of dimensions $\mathcal{A} = L_x \times L_y = (25 \times 25)\mu$m. The harmonic confinement along $z$ is sufficiently strong to freeze all degrees of freedom in that direction. The frequency of the harmonic potential, $\omega_z$, can be used to relate the actual 3D $s$-wave scattering length $a_{ij}$ of the atoms in different hyperfine levels to the 2D coupling constants $g$ and $g_{12}$ used in Eq.~(\ref{sgpe}), {\it via} the relation  $g_{ij} = \sqrt{8\pi} (\hbar^2/m) a_{ij}/a_z$, where $a_z = \sqrt{\hbar/m \omega_z}$ is the harmonic oscillator length. In our simulations, we use the mass of $^{87}$Rb atoms, the scattering length $a_{ii}=100 a_B$, where $a_B$ is the Bohr radius,  and $\omega_z= 2 \pi \times (1500\ {\rm Hz})$. A useful energy scale is given by the quantity $gn$, where $n$ is the total density of the gas. If the total number of atoms is $N=10^4$, then $gn=1.17 \times 10^{-31}$~J. For the coupling strength we use $\Omega = 0.1 gn$, which is in a range accessible to on-going experiments~\cite{farolfi_21,cominotti_21}. We also introduce the temperature $T_g=gn/k_B$, which will be used to present our results using  dimensionless units. }

\changes{We obtain the equilibrium configurations at a given  temperature $T$ and a given interaction strength $g_{12}$, by numerically propagating Eq.~(\ref{sgpe}) in real-time starting from purely random $c$-fields until equilibrium is reached. From now on, we consider the case of zero detuning ($\delta=0$). It is worth noticing that the total number of atoms $N$ is not an input of SGPE and it varies during each simulation; it stabilizes at final mean value, with tiny fluctuations, when the equilibrium configuration is reached. In fact, we use the stability of the mean value $N$ and the smallness of its fluctuations as criteria to stop each simulation. The input quantity is the chemical potential $\mu$, which is chosen in such a way that the equilibrium atom number $N$ is always very close to $N=10^4$, in all cases. The largest temperature in our simulations is $T/T_g=0.6$, which is about $0.15 T_{\rm BKT}$, where $T_{\rm BKT}$ is the critical temperature of the Kosterlitz-Berezinski-Thouless transition for a single component Bose gas of density $n$ in the same geometry~\cite{prokofev_01}. We are thus well inside the superfluid phase of the mixture. A consequence is that quantized vortices are absent in our configuration at equilibrium.} 

\changes{Typical trajectories of the magnetization $Z$ {\it vs.} time are shown Fig.~\ref{oscnum}(d) for simulations lasting $3$~s, at the same temperature $T/T_g=0.2$ but with different values of the interaction strength $g_{12}/{\bar g}_{12}$. The three panels (a)-(c) in the same figure are snapshots of the magnetization density profile, $(n_1-n_2)/(n_1+n_2)$, at the end of the simulation interval, i.e., in correspondence to the marker at the end of each trajectory in panel (d). For $g_{12}/{\bar g}_{12}=0.95$, which is well inside the paramagnetic phase, the magnetization density exhibits weak fluctuations  and the magnetization $Z$ remains always small during the evolution. For $g_{12}/{\bar g}_{12}=1.01$, fluctuations are larger and persist for long times. Finally, for $g_{12}/{\bar g}_{12}=1.02$, the gas quite rapidly polarizes and $Z$ shows small fluctuations around a finite value, as expected in the ferromagnetic phase. Note also that the phase transition is not accompanied by spatial separation; in fact, the number of atoms in each hyperfine state is not conserved and atoms can switch from one state to the other at any point is space, while the total number $N$ is conserved. This implies that, in the paramagnetic phase, the gas exhibits a randomly fluctuating population imbalance, with zero mean magnetization, while in the ferromagnetic phase the atoms prefer to occupy the same (randomly chosen) state in the whole volume. This is different from the case of two uncoupled condensates, each one composed by a fixed fraction of atoms, where a spatial separation occurs for $g_{12}>g$, corresponding to a miscible-immiscible transition. }
 
\changes{The phase transition at each temperature $T$ can be characterized by computing the quantity ${\cal Z}$ defined as the modulus of the averaged magnetization, 
\begin{equation}
{\cal Z} = | \langle Z \rangle | = \bigg\vert \bigg\langle \frac{N_1-N_2}{N_1+N_2} \bigg\rangle \bigg\vert \ ,
\label{posZ}
\end{equation} 
and the variance 
\begin{equation}
(\Delta Z)^2 = \bigg\langle \bigg(\frac{N_1-N_2}{N_1+N_2}\bigg)^2 \bigg\rangle - \bigg(\bigg\langle \frac{N_1-N_2}{N_1+N_2} \bigg\rangle \bigg)^2 \, .
\end{equation} 
These averages include a time-average and a configuration-average. In particular, for each trajectory of the magnetization $Z$ {\it vs.} time, as those in Fig.~\ref{oscnum}(d), a time average is carried out over a time interval when the gas is at equilibrium and the mean value of the magnetization is sufficiently stable. Furthermore, an ensemble average is performed over a large number (${\mathcal N} \approx 500$) of time-averaged trajectories in order to suppress the effects of random noise. As a result, the residual uncertainty on the values of ${\cal Z}$ and $(\Delta Z)^2$ is drastically reduced, in such a way that we can plot the SGPE data without error bars, given that the statistical errors are of the same order of the marker size in figures of the next section. }

\begin{figure*}[!hbtp]
\centering
\includegraphics[width=\textwidth]{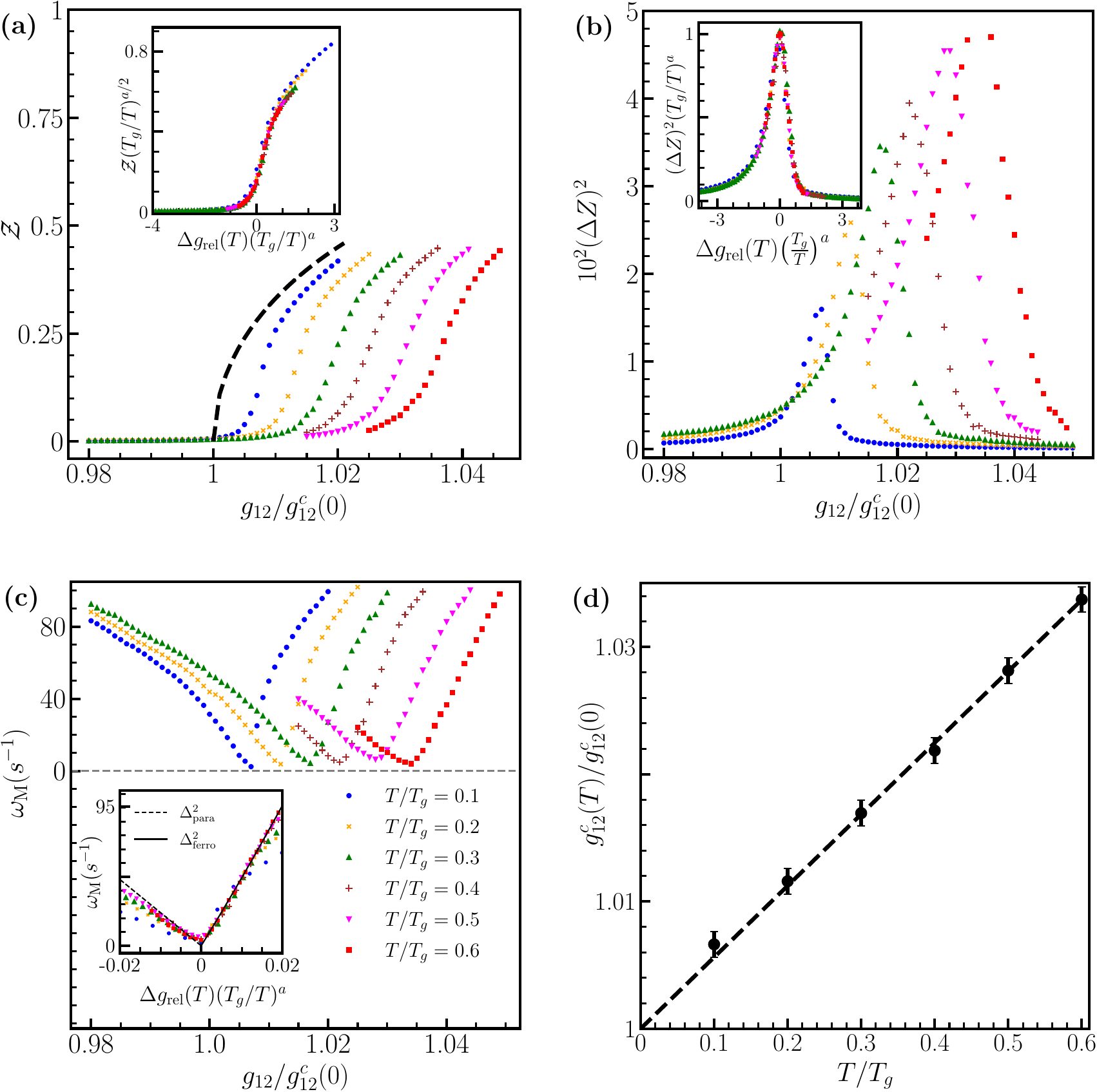}
\caption{Results of SGPE simulations at finite temperature and zero detuning for: (a) the average magnetization ${\cal Z}$ (see Eq.~(\ref{posZ})); (b) its variance $(\Delta Z)^2$; (c) the dominant relaxation frequency $\omega_{\rm M}$. In the main plots, these quantities are given as a function of the interaction parameter $g_{12}/g_{12}^c(0)$, where $g_{12}^c(0)$ is the critical value at $T=0$ defined in Eq.~(\ref{eq:criticalp}). The correspondence between temperature, in units of $T_g=gn/k_B$, and type of markers is given in the legend in panel (c). The black dashed line in (a) is the $T=0$ mean-field magnetization obtained from Eq.~(\ref{mfmag}) for a homogeneous gas. Panel (d) shows the variation of the critical interaction parameter $g_{12}^c(T)$ with temperature, as extracted from the location of the maxima of $(\Delta Z)^2$ in (b) and the minima of $\omega_{\rm M}$ in (c); the difference between the two estimates is smaller than the error bar, which is identified as the grid spacing in the parameter space of SGPE simulations; the dashed line is a linear fit.
The inset in (a) shows the magnetization vs. the relative distance from the critical point, defined in Eq.~(\ref{eq:reldist}), with a power-law rescaling as in (\ref{eq:scalinz}), with $a=3/5$, producing a collapse of all points on a single curve in the critical region; in the inset of panels (b) and (c), a similar rescaling is applied to $(\Delta Z)^2$, according to Eq.~(\ref{eq:scalindeltaz}), and to  $\omega_{\rm M}$, respectively, with the same exponent $a$. The dashed line in the inset of panel (c) corresponds to the $\Delta_{_{\rm para}}^2$, Eq.~(\ref{eq:gappara}), and the solid line on the right corresponds to $\Delta_{_{\rm ferro}}^2$, Eq.~(\ref{eq:gapferro}).
}
\label{flucmag}
\end{figure*}

\section{Equilibrium properties around the critical point}

\subsection{Shift of the critical point with temperature}

Our main results are reported in Fig.~\ref{flucmag}. The magnetization ${\cal Z}$ and the variance $(\Delta Z)^2$ are presented in panels (a) and (b), respectively, as a function of  $g_{12}/g_{12}^c(0)$, where 
\begin{equation}
g_{12}^c(0)={\bar g}_{12} = g + 2\Omega/n
\label{eq:criticalp}
\end{equation}
is the critical point at $T=0$ and in panel (a), we also report the magnetization at zero temperature in the ferromagnetic phase of a homogeneous gas, obtained from Eq.~(\ref{mfmag}). Different colors and markers refer to simulations at different temperature. The lowest value that we consider is $T/T_g=0.1$ and corresponds to the blue circles, while the highest temperature is $T/T_g=0.6$, corresponding to the red squares, as indicated in the legend in panel (c). For each temperature, the para-ferromagnetic transition is signaled by a rapid increase of magnetization ${\cal Z}$ and a maximum of its variance $(\Delta Z)^2$.  A high variance in magnetization implies large fluctuations from the mean, which indeed happens at the ferromagnetic critical point. 
The emergence of global magnetization fluctuations at the transition can be detected in  experiments~\cite{kristensen_19,christensen_21}.

The results suggest that the critical interaction parameter $g_{12}^c(T)$ is linearly shifted upwards with $T$. 

In order to obtain an accurate estimate $g_{12}^c(T)$, we also use a third indicator, namely the frequency characterising the thermalization of the magnetization, a quantity which is expected to show a critical slowing down close to the phase transition. Indeed when the system approaches the critical point it equilibrates more slowly. In particular, each trajectory in the SGPE simulations tends to exhibit oscillations of the magnetization around the equilibrium value, with a period which is larger close to the critical point, as shown in  Fig.~\ref{oscnum}(c). We calculate the Fourier transform of the magnetization for each trajectory and extract the dominant relaxation frequency, $\omega_{\rm M}$, and then we perform an ensemble average. The resulting values of the frequency $\omega_{\rm M}$ are reported in panel (c) of the Fig.~\ref{flucmag}. As expected, we find a strong decrease of dominant frequency at the transition, corresponding to a large increase of the equilibration time. 

The transition point is located at value of $g_{12}/g_{12}^c(0)$ where $(\Delta Z)^2$ is maximum and $\omega_{\rm M}$ is minimum. We determine the positions of the maxima in Fig.~\ref{flucmag}(b) and minima in Fig.~\ref{flucmag}(c) with a quadratic fit to the closest points. The two estimates almost coincide. In Fig.~\ref{flucmag}(d), we plot the critical points $g_{12}^c(T)/g_{12}^c(0)$ obtained as the average of the two estimates for each $T/T_g$. The error bars are simply the grid spacing in the parameter space of our simulations; the distance between the location of the maximum in (b) and minimum in (c) for each $T/T_g$ is less than the error bar. The critical value $g_{12}^c(T)$ turns out to increase linearly with $T$; the dashed line is a linear fit to the data. The slope is such that the shift of the critical point at finite temperature remains relatively small in the range of $T$ here considered. 

We can relate our findings in Fig.~\ref{flucmag}(d) to the so called ``shift critical exponent", $\Psi$, which identifies the critical line at finite temperature close to a quantum critical point \cite{Continentino}. This can be done by writing $g^c_{12}(T)=g^c_{12}(0)(1+uT^{1/\Psi})$. Our results suggest that the shift critical exponent is $\Psi=1$, with $u\simeq 0.056$,  even for relatively large temperature, where still the deviation from $g^c_{12}(0)$ is small due to the smallness of the prefactor $u$. 

\begin{figure}[!hbtp]
\centering
\includegraphics[width=0.95\columnwidth]{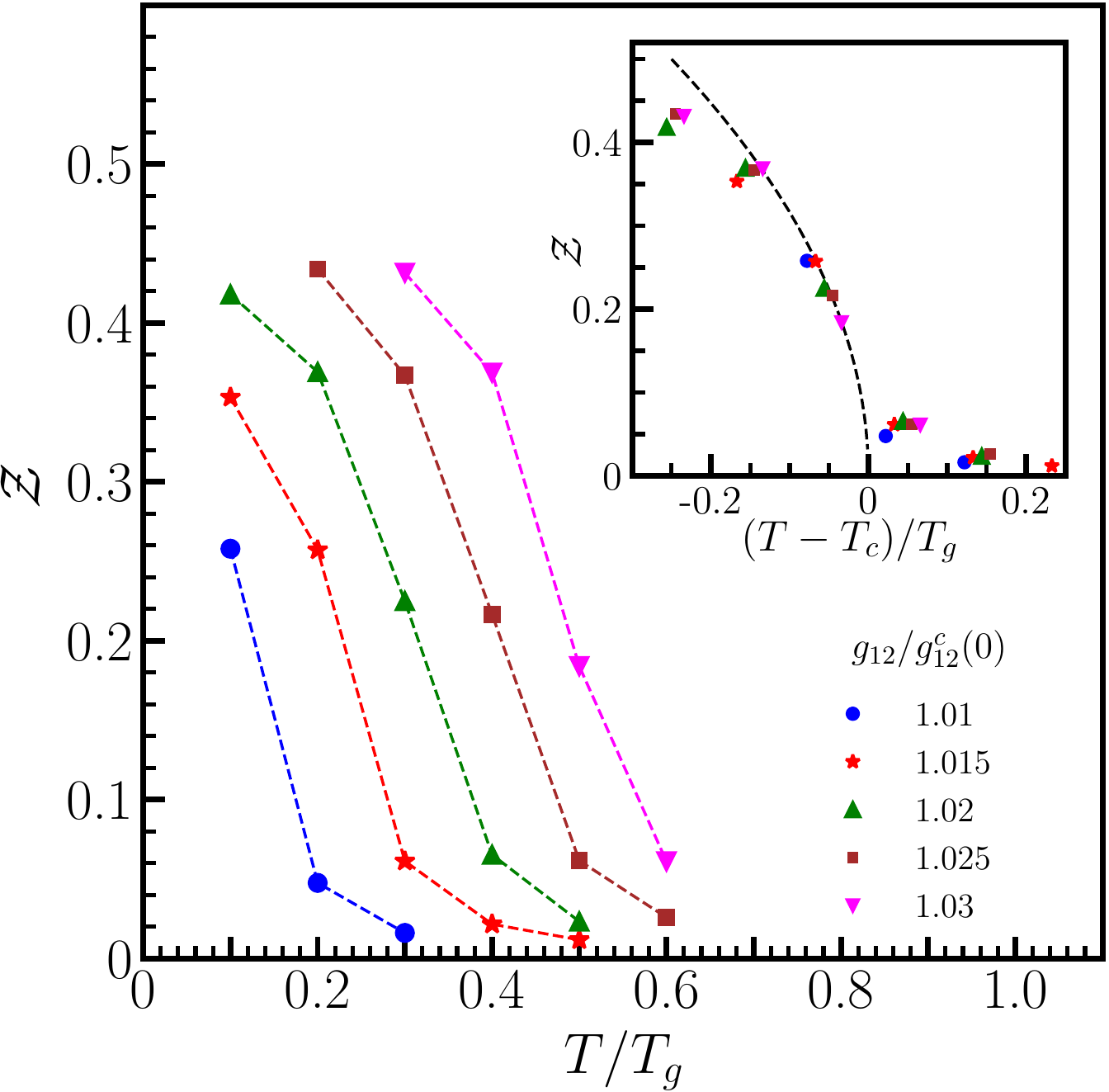}
\caption{Average magnetization ${\cal Z}$ (see Eq.~(\ref{posZ})) as a function of $T/T_g$ for fixed values of  $g_{12}/g_{12}^c(0)$. Points correspond to the results of SGPE as in Fig.~\ref{flucmag}(a), while lines are guide to the eyes. The same data are shown in the inset, but plotted as a function of the relative distance from the critical temperature, $\delta T = (T-T_c)/T_g$, where $T_c$ is extracted from the linear fit to the critical points in  Fig.~\ref{flucmag}(d). The dashed line represents the power-law $|\delta T|^{1/2}$. }
\label{mag-varT}
\end{figure}

Finally, in Fig. \ref{mag-varT} we show the magnetization ${\cal Z}$ as a function of $T/T_g$ for a set of different values of $g_{12}$. As shown in the inset, the magnetization bears a thermal power-law scaling behaviour with the relative temperature, $\delta T = (T-T_c)/T_g$, given by ${\cal Z}=|\delta T|^{1/2}$, i.e., it presents a thermal mean-field critical exponent $\beta=1/2$. \changes{The dashed line is expected to be valid only for small $\delta T$; the fact that the SGPE results for ${\cal Z}$ remain finite on the right of the critical point is consistent with the observation of strong fluctuations (large  $(\Delta Z)^2$) around the critical temperature and with finite size effects. }

\subsection{Universal scaling around the critical region}

In the following we show that the ${\cal Z}$, $(\Delta Z)^2$ and $\omega_{\rm M}$ exhibit nice scaling properties with the temperature. For this purpose, we first define the relative distance from the critical point as 
\begin{equation}
\Delta g_{\rm rel}(T) = \frac{g_{12} - g_{12}^c(T)}{g_{12}^c(0)} \ ,
\label{eq:reldist}
\end{equation} 
and we use it to shift the SGPE results for all quantities in 
Fig.~\ref{flucmag}(a)-(c).

\paragraph{Magnetic fluctuations.}
The scaling can be better appreciated for the magnetic fluctuation curves, given their smooth shape. 
Indeed we find that both the FWHM and the peak height of $(\Delta Z)^2$ in Fig.~\ref{flucmag}(b) vary with temperature as $(T/T_g)^a$, where the exponent can be calculated by fitting the two quantities with 
a power law; the results are $a=0.63 \pm 0.05$ and $a=0.57 \pm 0.05$, respectively. For simplicity, and remaining with the error bars, we choose 
the exponent to be the same and rescale $(\Delta Z)^2$ in the form 
\begin{equation} 
(\Delta Z)^2 (\Delta g_{\rm rel}(T),T)= T^{a} F (\Delta g_{\rm rel}(T)/T^a) \, ,
\label{eq:scalindeltaz}
\end{equation}
with $a$ equal to $3/5$. As shown in the inset of Fig.~\ref{flucmag}(b), 
all points of the SGPE simulations nicely collapse onto a single universal 
curve, in agreement with the above scaling law, in the whole range of temperature here considered.

\paragraph{Magnetization.}
We find that a similar scaling behavior applies to the magnetization ${\cal Z}$. 
In particular, as one can see in the inset of Fig.~\ref{flucmag}(a), the scaling works well with
\begin{equation} 
{\cal Z} (\Delta g_{\rm rel}(T),T)= T^{a/2} F (\Delta g_{\rm rel}(T)/T^a) \, ,
\label{eq:scalinz}
\end{equation}
where $a$ is the same as before. The factor $a/2$ in the first exponent is consistent with an extrapolation to $T=0$, where the SGPE is expected to reproduce the mean-field prediction ${\cal Z}\propto [\Delta g_{\rm rel}(0)]^{1/2}$; this $T=0$ prediction is represented by the dashed line in Fig.~\ref{flucmag}(a) for the case of a uniform gas in the thermodynamic limit.  

\paragraph{Critical slowing down.} 
The relaxation frequency $\omega_{\rm M}$, related to the critical slowing down, provides an insight in the relaxation dynamics near the phase transition. If we plot $\omega_{\rm M}$ as a function of the rescaled variable $\Delta g_{\rm rel}(T)(T_g/T)^a$, as we did for ${\cal Z}$ and $(\Delta Z)^2$, with the same $a$, again all SGPE results exhibit a reasonably good collapse onto a universal curve, as shown in the inset of Fig.~\ref{flucmag}(c). Furthermore, one can observe that the relaxation frequency scales as $\omega_{\rm M}\propto \Delta g_{\rm rel}$, both on the right and the left of the transition, but with a different slope, namely two times larger in the ferromagnetic phase than in the paramagnetic.
According to the general definition of the dynamical critical exponents \cite{hohenberg_77} this behaviour would be consistent with a value $\nu z=1$. 

The critical slowing is related to the divergence of the susceptibility of the system, which is due to the closure of the excitation gap at the critical point. At $T=0$ -- unlike the uncoupled Bose-Bose mixtures -- the coherently coupled gas has a spin gap in both the paramagnetic and ferromagnetic phases given by \cite{recati_21,notefactor2}
\begin{eqnarray}
\Delta_{\rm para}& =&\sqrt{2\Omega[(g-g_{12})n + 2\Omega]} \\
\Delta_{\rm ferro}& =& \sqrt{[(g-g_{12})n]^2 - (2\Omega)^2]} \ ,
\end{eqnarray}
respectively. Close to the critical point, these expressions satisfy the general relation $\Delta_{\rm ferro}=\sqrt{2}\Delta_{\rm para}$, which is characteristic of a $\mathbb{Z}_2$ phase transition. The square of the previous expressions as a function of $\Delta g_{\rm rel}$ at $T=0$ read
\begin{eqnarray}
\Delta^2_{\rm para}(\Delta g_{\rm rel})\!\! & = & 2 n \Omega g^c_{12} |\Delta g_{\rm rel}| 
\label{eq:gappara} \\
\Delta^2_{\rm ferro}(\Delta g_{\rm rel})\!\! & = & 4n \Omega g^c_{12}\Delta g_{\rm rel} \left(1 + \frac{n g^c_{12} \Delta g_{\rm rel}}{\Omega} \right) . 
\label{eq:gapferro}
\end{eqnarray}
This suggests a proportionality between the relaxation frequency and the square of the gap. Interestingly, within our SGPE approach, we find that the relaxation frequency as a function of $x=\Delta g_{\rm{rel}}(T)(T_g/T)^a$, is very well approximated just by $\Delta^2$, i.e,   
\begin{eqnarray}
\omega_{\rm M}(x)&\simeq&\Delta^2_{\rm para}(x) \; \mathrm{for}\; x<1 
\label{eq:gappara} \\
\omega_{\rm M}(x)&\simeq&\Delta^2_{\rm ferro}(x)  \; \mathrm{for}\; x>1 \ . 
\label{eq:gapferro}
\end{eqnarray}
as shown by the black lines in the inset of Fig.~\ref{flucmag}(c). 

While our numerical SPGE results present an overall consistency in their behavior across the transition, based on general arguments on the role of the critical exponents, it is worth stressing that the exponent $a=3/5$ entering the scaling functions should be taken as a purely numerical outcome, not as an exact value. In fact, here we do not pretend to extract the scaling exponents with high precision, but rather to provide a first quantitative characterisation of the finite temperature para- to ferro-magnetic transition in a spinor superfluid in terms of plausible scaling behaviors, using a theory which is known to account for thermal fluctuations to a good level of approximation.  

\begin{figure}[!hbtp]
\centering
\includegraphics[width=0.95\columnwidth]{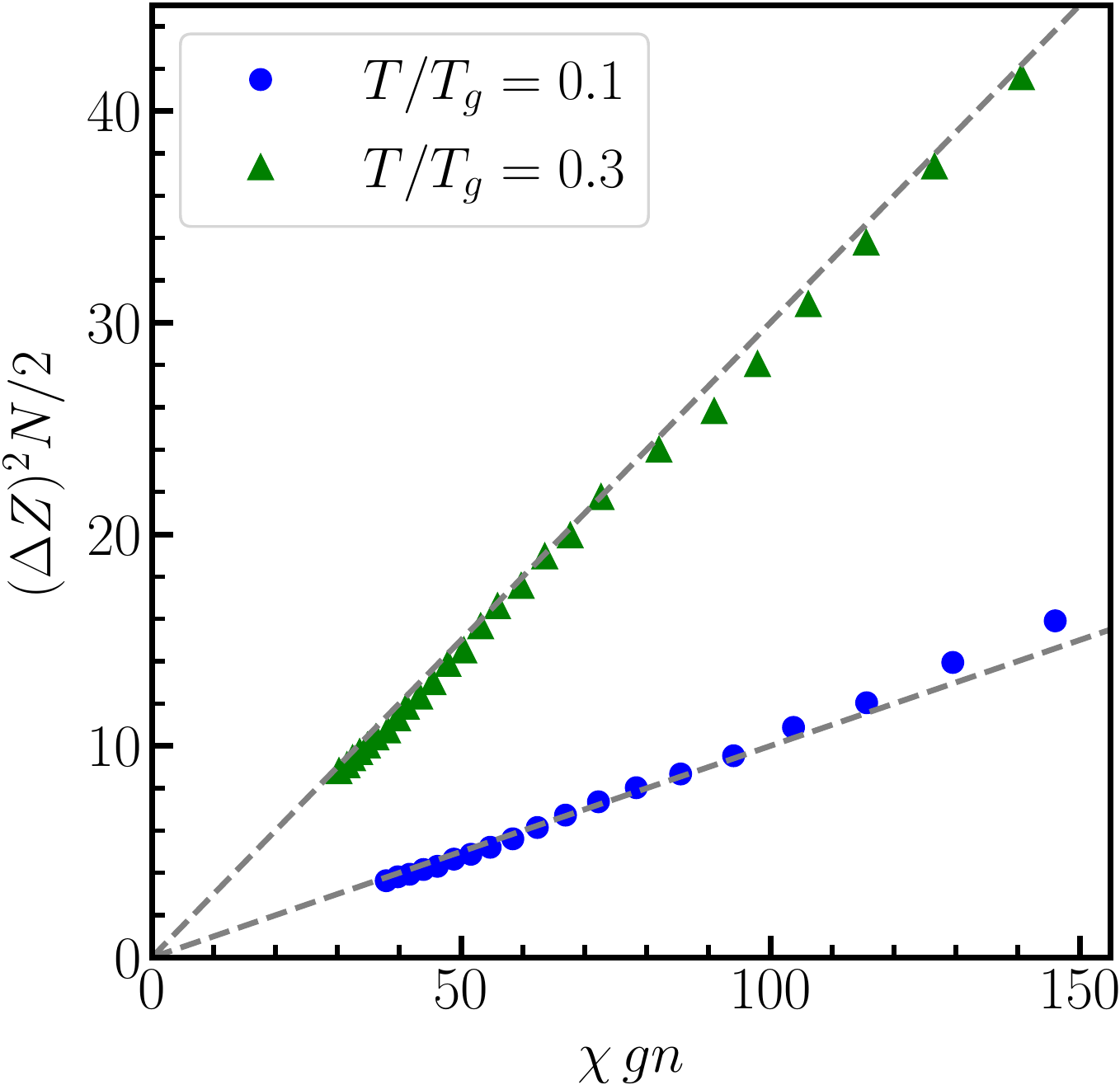}
\caption{Relation between $(\Delta Z)^2$ and the magnetic susceptibility $\chi$; dashed lines represent the prediction of the fluctuation-dissipation theorem according to Eq.~(\ref{eq:fluctdiss}).}
\label{fluctuation-dissipation}
\end{figure}

\subsection{Fluctuation-dissipation theorem}

As a final test of our SGPE results, we check the applicability of the fluctuation-dissipation theorem for the spin channel.

\changes{The fluctuations of the order parameter  are related to the excitation spectrum of the system via the relation $\Delta Z^2= S_s(0,T)$, where $S_s(q,T)$ is the spin static structure factor. 
The latter can be easily evaluated within Bogoliubov theory for coherently coupled gases \cite{abad_13} and for $q=0$ it reads 
\begin{equation}
S_s(0,T)=\sqrt{\frac{\Omega\chi}{4}}\coth\left(\frac{\sqrt{2\Omega\chi^{-1}}}{k_B T}\right),
\end{equation}  
where $\chi$ is the magnetic susceptibility, defined as 
 $\chi = \lim_{\delta\rightarrow 0}d{\cal Z}/d\delta$ using Eq.~(\ref{sgpe}).
In particular, close to the critical point, where the temperature is the dominant energy scale (larger than the spin gap),} the fluctuation-dissipation relation takes the classical form $N(\Delta Z)^2 = 2 k_{\rm B} T \chi$, or 
\begin{equation}
\frac{N}{2}(\Delta Z)^2  = \frac{T}{T_g} \chi gn \, .
\label{eq:fluctdiss}
\end{equation}

The calculation of $\chi$ through SGPE demands heavy computational efforts and we have
restricted ourselves to the paramagnetic phase and only two temperatures, namely $T/T_g = 0.1$ and $0.3$, and not too close to the critical point, where the numerical calculation of derivates becomes unreliable.
For fixed $T/T_g$, Eq.~(\ref{eq:fluctdiss}) predicts a linear relation between $(\Delta Z)^2 N/2$ and $\chi gn$. In Fig.~\ref{fluctuation-dissipation}, we show the straight lines representing Eq.~(\ref{eq:fluctdiss}) together with the results obtained from SGPE simulations, with $k_{\rm B}T/\Delta_{\rm para}$ ranging from 2 to 20. The good agreement demonstrates that the simulations accurately account for the fluctuation-dissipation theorem.

\section{Conclusions}

In this work, we have investigated the finite temperature paramagnetic to ferromagnetic transition in coherently coupled weakly interacting Bose-Einstein condensates in two-dimensions, by using the Stochastic (projected) Gross-Pitaevskii theory. 
Marked by a sharp increase of the average magnetization, enhanced magnetic fluctuations, and a strong increase of the relaxation time, the transition is found to occur along a critical line corresponding to a linear shift of the quantum critical point with temperature. The fluctuations of the magnetization are shown to exhibit a linear relationship with the spin-susceptibility in agreement with the fluctuation-dissipation theorem. The SGPE results for the magnetization, the magnetic fluctuations and the relaxation frequency (critical slowing down) turn out to collapse onto universal curves upon a proper rescaling of the critical point with temperature. Moreover, close to the transition the relaxation frequency appears to simply coincide with the square of the spin excitation gap. Given that the SGPE simulations are rather time consuming, we have so far restricted the analysis to a single square box of size comparable to that of available two-dimensional box-like traps of current experiments; further calculations with larger boxes would be needed for a more accurate determination of the scaling exponents in the thermodynamic limit.

%%%%%%%%%%%%%%%%%%%%%%%%%%%%%%%%%%%%%%%%%%%%%%%%%%%%%%%%%%%%%%%%%%%%%%%%%%%%%
%%%%%%%%%%%%%%%%%%             Acknowledgements             %%%%%%%%%%%%%%%%%
%%%%%%%%%%%%%%%%%%%%%%%%%%%%%%%%%%%%%%%%%%%%%%%%%%%%%%%%%%%%%%%%%%%%%%%%%%%%%
\begin{acknowledgments}
 This work is supported by Provincia autonoma di Trento and from INFN-TIFPA under the project FIS$\hbar$. We acknowledge the CINECA award under the ISCRA initiative, for the availability of high performance computing resources and support. A.Roy acknowledges the support of the Science and Engineering Research Board (SERB), Department of Science and Technology, Government of 
 India under the project SRG/2022/000057 and IIT Mandi seed-grant funds under the project IITM/SG/AR/87. A.Roy acknowledges National Supercomputing Mission (NSM) for providing computing resources of PARAM Himalaya at IIT Mandi, which is implemented by C-DAC and supported by the Ministry of Electronics and
Information Technology (MeitY) and Department of Science and Technology (DST), Government of India.
 
\end{acknowledgments}

%%%%%%%%%%%%%%%%%%%%%%%%%%%%%%%%%%%%%%%%%%%%%%%%%%%%%%%%%%%%%%%%%%%%%%%%%%%%%
%%%%%%%%%%%%%%%%%              Bibliography                  %%%%%%%%%%%%%%%%
%%%%%%%%%%%%%%%%%%%%%%%%%%%%%%%%%%%%%%%%%%%%%%%%%%%%%%%%%%%%%%%%%%%%%%%%%%%%%
\bibliography{refs}{}

%merlin.mbs apsrev4-1.bst 2010-07-25 4.21a (PWD, AO, DPC) hacked
%Control: key (0)
%Control: author (72) initials jnrlst
%Control: editor formatted (1) identically to author
%Control: production of article title (-1) disabled
%Control: page (0) single
%Control: year (1) truncated
%Control: production of eprint (0) enabled
\begin{thebibliography}{101}%
\makeatletter
\providecommand \@ifxundefined [1]{%
 \@ifx{#1\undefined}
}%
\providecommand \@ifnum [1]{%
 \ifnum #1\expandafter \@firstoftwo
 \else \expandafter \@secondoftwo
 \fi
}%
\providecommand \@ifx [1]{%
 \ifx #1\expandafter \@firstoftwo
 \else \expandafter \@secondoftwo
 \fi
}%
\providecommand \natexlab [1]{#1}%
\providecommand \enquote  [1]{``#1''}%
\providecommand \bibnamefont  [1]{#1}%
\providecommand \bibfnamefont [1]{#1}%
\providecommand \citenamefont [1]{#1}%
\providecommand \href@noop [0]{\@secondoftwo}%
\providecommand \href [0]{\begingroup \@sanitize@url \@href}%
\providecommand \@href[1]{\@@startlink{#1}\@@href}%
\providecommand \@@href[1]{\endgroup#1\@@endlink}%
\providecommand \@sanitize@url [0]{\catcode `\\12\catcode `\$12\catcode
  `\&12\catcode `\#12\catcode `\^12\catcode `\_12\catcode `\%12\relax}%
\providecommand \@@startlink[1]{}%
\providecommand \@@endlink[0]{}%
\providecommand \url  [0]{\begingroup\@sanitize@url \@url }%
\providecommand \@url [1]{\endgroup\@href {#1}{\urlprefix }}%
\providecommand \urlprefix  [0]{URL }%
\providecommand \Eprint [0]{\href }%
\providecommand \doibase [0]{http://dx.doi.org/}%
\providecommand \selectlanguage [0]{\@gobble}%
\providecommand \bibinfo  [0]{\@secondoftwo}%
\providecommand \bibfield  [0]{\@secondoftwo}%
\providecommand \translation [1]{[#1]}%
\providecommand \BibitemOpen [0]{}%
\providecommand \bibitemStop [0]{}%
\providecommand \bibitemNoStop [0]{.\EOS\space}%
\providecommand \EOS [0]{\spacefactor3000\relax}%
\providecommand \BibitemShut  [1]{\csname bibitem#1\endcsname}%
\let\auto@bib@innerbib\@empty
%</preamble>
\bibitem [{\citenamefont {Stamper-Kurn}\ and\ \citenamefont
  {Ueda}(2013)}]{stamperkurn_13}%
  \BibitemOpen
  \bibfield  {author} {\bibinfo {author} {\bibfnamefont {D.~M.}\ \bibnamefont
  {Stamper-Kurn}}\ and\ \bibinfo {author} {\bibfnamefont {M.}~\bibnamefont
  {Ueda}},\ }\href {\doibase 10.1103/RevModPhys.85.1191} {\bibfield  {journal}
  {\bibinfo  {journal} {Rev. Mod. Phys.}\ }\textbf {\bibinfo {volume} {85}},\
  \bibinfo {pages} {1191} (\bibinfo {year} {2013})}\BibitemShut {NoStop}%
\bibitem [{\citenamefont {Pethick}\ and\ \citenamefont
  {Smith}(2008)}]{pethick_08}%
  \BibitemOpen
  \bibfield  {author} {\bibinfo {author} {\bibfnamefont {C.}~\bibnamefont
  {Pethick}}\ and\ \bibinfo {author} {\bibfnamefont {H.}~\bibnamefont
  {Smith}},\ }\href {http://books.google.co.in/books?id=HobTUdxBoFcC} {\emph
  {\bibinfo {title} {{B}ose-{E}instein Condensation in Dilute Gases}}}\
  (\bibinfo  {publisher} {Cambridge University Press, New York},\ \bibinfo
  {year} {2008})\BibitemShut {NoStop}%
\bibitem [{\citenamefont {Kawaguchi}\ and\ \citenamefont
  {Ueda}(2012{\natexlab{a}})}]{kawaguchi}%
  \BibitemOpen
  \bibfield  {author} {\bibinfo {author} {\bibfnamefont {Y.}~\bibnamefont
  {Kawaguchi}}\ and\ \bibinfo {author} {\bibfnamefont {M.}~\bibnamefont
  {Ueda}},\ }\href {\doibase https://doi.org/10.1016/j.physrep.2012.07.005}
  {\bibfield  {journal} {\bibinfo  {journal} {Physics Reports}\ }\textbf
  {\bibinfo {volume} {520}},\ \bibinfo {pages} {253} (\bibinfo {year}
  {2012}{\natexlab{a}})}\BibitemShut {NoStop}%
\bibitem [{\citenamefont {Pitaevskii}\ and\ \citenamefont
  {Stringari}(2016{\natexlab{a}})}]{pitaevskii-16}%
  \BibitemOpen
  \bibfield  {author} {\bibinfo {author} {\bibfnamefont {L.}~\bibnamefont
  {Pitaevskii}}\ and\ \bibinfo {author} {\bibfnamefont {S.}~\bibnamefont
  {Stringari}},\ }\href@noop {} {\emph {\bibinfo {title} {{B}ose-{E}instein
  Condensation}}},\ International Series of Monographs on Physics\ (\bibinfo
  {publisher} {Clarendon Press, U. K.},\ \bibinfo {year} {2016})\BibitemShut
  {NoStop}%
\bibitem [{\citenamefont {Ho}\ and\ \citenamefont {Shenoy}(1996)}]{ho_96}%
  \BibitemOpen
  \bibfield  {author} {\bibinfo {author} {\bibfnamefont {T.-L.}\ \bibnamefont
  {Ho}}\ and\ \bibinfo {author} {\bibfnamefont {V.~B.}\ \bibnamefont
  {Shenoy}},\ }\href {\doibase 10.1103/PhysRevLett.77.3276} {\bibfield
  {journal} {\bibinfo  {journal} {Phys. Rev. Lett.}\ }\textbf {\bibinfo
  {volume} {77}},\ \bibinfo {pages} {3276} (\bibinfo {year}
  {1996})}\BibitemShut {NoStop}%
\bibitem [{\citenamefont {Timmermans}(1998)}]{timmermans_98}%
  \BibitemOpen
  \bibfield  {author} {\bibinfo {author} {\bibfnamefont {E.}~\bibnamefont
  {Timmermans}},\ }\href {\doibase 10.1103/PhysRevLett.81.5718} {\bibfield
  {journal} {\bibinfo  {journal} {Phys. Rev. Lett.}\ }\textbf {\bibinfo
  {volume} {81}},\ \bibinfo {pages} {5718} (\bibinfo {year}
  {1998})}\BibitemShut {NoStop}%
\bibitem [{\citenamefont {Ao}\ and\ \citenamefont {Chui}(1998)}]{ao_98}%
  \BibitemOpen
  \bibfield  {author} {\bibinfo {author} {\bibfnamefont {P.}~\bibnamefont
  {Ao}}\ and\ \bibinfo {author} {\bibfnamefont {S.~T.}\ \bibnamefont {Chui}},\
  }\href {\doibase 10.1103/PhysRevA.58.4836} {\bibfield  {journal} {\bibinfo
  {journal} {Phys. Rev. A}\ }\textbf {\bibinfo {volume} {58}},\ \bibinfo
  {pages} {4836} (\bibinfo {year} {1998})}\BibitemShut {NoStop}%
\bibitem [{\citenamefont {Trippenbach}\ \emph {et~al.}(2000)\citenamefont
  {Trippenbach}, \citenamefont {G\'oral}, \citenamefont {Rzazewski},
  \citenamefont {Malomed},\ and\ \citenamefont {Band}}]{trippenbach_2000}%
  \BibitemOpen
  \bibfield  {author} {\bibinfo {author} {\bibfnamefont {M.}~\bibnamefont
  {Trippenbach}}, \bibinfo {author} {\bibfnamefont {K.}~\bibnamefont
  {G\'oral}}, \bibinfo {author} {\bibfnamefont {K.}~\bibnamefont {Rzazewski}},
  \bibinfo {author} {\bibfnamefont {B.}~\bibnamefont {Malomed}}, \ and\
  \bibinfo {author} {\bibfnamefont {Y.~B.}\ \bibnamefont {Band}},\ }\href
  {http://stacks.iop.org/0953-4075/33/i=19/a=314} {\bibfield  {journal}
  {\bibinfo  {journal} {J. Phys. B}\ }\textbf {\bibinfo {volume} {33}},\
  \bibinfo {pages} {4017} (\bibinfo {year} {2000})}\BibitemShut {NoStop}%
\bibitem [{\citenamefont {Van~Schaeybroeck}(2008)}]{schaeybroeck_08}%
  \BibitemOpen
  \bibfield  {author} {\bibinfo {author} {\bibfnamefont {B.}~\bibnamefont
  {Van~Schaeybroeck}},\ }\href {\doibase 10.1103/PhysRevA.78.023624} {\bibfield
   {journal} {\bibinfo  {journal} {Phys. Rev. A}\ }\textbf {\bibinfo {volume}
  {78}},\ \bibinfo {pages} {023624} (\bibinfo {year} {2008})}\BibitemShut
  {NoStop}%
\bibitem [{\citenamefont {Wen}\ \emph {et~al.}(2012)\citenamefont {Wen},
  \citenamefont {Liu}, \citenamefont {Cai}, \citenamefont {Zhang},\ and\
  \citenamefont {Hu}}]{wen_12}%
  \BibitemOpen
  \bibfield  {author} {\bibinfo {author} {\bibfnamefont {L.}~\bibnamefont
  {Wen}}, \bibinfo {author} {\bibfnamefont {W.~M.}\ \bibnamefont {Liu}},
  \bibinfo {author} {\bibfnamefont {Y.}~\bibnamefont {Cai}}, \bibinfo {author}
  {\bibfnamefont {J.~M.}\ \bibnamefont {Zhang}}, \ and\ \bibinfo {author}
  {\bibfnamefont {J.}~\bibnamefont {Hu}},\ }\href {\doibase
  10.1103/PhysRevA.85.043602} {\bibfield  {journal} {\bibinfo  {journal} {Phys.
  Rev. A}\ }\textbf {\bibinfo {volume} {85}},\ \bibinfo {pages} {043602}
  (\bibinfo {year} {2012})}\BibitemShut {NoStop}%
\bibitem [{\citenamefont {Modugno}\ \emph {et~al.}(2002)\citenamefont
  {Modugno}, \citenamefont {Modugno}, \citenamefont {Riboli}, \citenamefont
  {Roati},\ and\ \citenamefont {Inguscio}}]{modugno_02}%
  \BibitemOpen
  \bibfield  {author} {\bibinfo {author} {\bibfnamefont {G.}~\bibnamefont
  {Modugno}}, \bibinfo {author} {\bibfnamefont {M.}~\bibnamefont {Modugno}},
  \bibinfo {author} {\bibfnamefont {F.}~\bibnamefont {Riboli}}, \bibinfo
  {author} {\bibfnamefont {G.}~\bibnamefont {Roati}}, \ and\ \bibinfo {author}
  {\bibfnamefont {M.}~\bibnamefont {Inguscio}},\ }\href {\doibase
  10.1103/PhysRevLett.89.190404} {\bibfield  {journal} {\bibinfo  {journal}
  {Phys. Rev. Lett.}\ }\textbf {\bibinfo {volume} {89}},\ \bibinfo {pages}
  {190404} (\bibinfo {year} {2002})}\BibitemShut {NoStop}%
\bibitem [{\citenamefont {Thalhammer}\ \emph {et~al.}(2008)\citenamefont
  {Thalhammer}, \citenamefont {Barontini}, \citenamefont {De~Sarlo},
  \citenamefont {Catani}, \citenamefont {Minardi},\ and\ \citenamefont
  {Inguscio}}]{thalhammer_08}%
  \BibitemOpen
  \bibfield  {author} {\bibinfo {author} {\bibfnamefont {G.}~\bibnamefont
  {Thalhammer}}, \bibinfo {author} {\bibfnamefont {G.}~\bibnamefont
  {Barontini}}, \bibinfo {author} {\bibfnamefont {L.}~\bibnamefont {De~Sarlo}},
  \bibinfo {author} {\bibfnamefont {J.}~\bibnamefont {Catani}}, \bibinfo
  {author} {\bibfnamefont {F.}~\bibnamefont {Minardi}}, \ and\ \bibinfo
  {author} {\bibfnamefont {M.}~\bibnamefont {Inguscio}},\ }\href {\doibase
  10.1103/PhysRevLett.100.210402} {\bibfield  {journal} {\bibinfo  {journal}
  {Phys. Rev. Lett.}\ }\textbf {\bibinfo {volume} {100}},\ \bibinfo {pages}
  {210402} (\bibinfo {year} {2008})}\BibitemShut {NoStop}%
\bibitem [{\citenamefont {Lercher}\ \emph {et~al.}(2011)\citenamefont
  {Lercher}, \citenamefont {Takekoshi}, \citenamefont {Debatin}, \citenamefont
  {Schuster}, \citenamefont {Rameshan}, \citenamefont {Ferlaino}, \citenamefont
  {Grimm},\ and\ \citenamefont {N\"agerl}}]{lercher_11}%
  \BibitemOpen
  \bibfield  {author} {\bibinfo {author} {\bibfnamefont {A.}~\bibnamefont
  {Lercher}}, \bibinfo {author} {\bibfnamefont {T.}~\bibnamefont {Takekoshi}},
  \bibinfo {author} {\bibfnamefont {M.}~\bibnamefont {Debatin}}, \bibinfo
  {author} {\bibfnamefont {B.}~\bibnamefont {Schuster}}, \bibinfo {author}
  {\bibfnamefont {R.}~\bibnamefont {Rameshan}}, \bibinfo {author}
  {\bibfnamefont {F.}~\bibnamefont {Ferlaino}}, \bibinfo {author}
  {\bibfnamefont {R.}~\bibnamefont {Grimm}}, \ and\ \bibinfo {author}
  {\bibfnamefont {H.-C.}\ \bibnamefont {N\"agerl}},\ }\href {\doibase
  10.1140/epjd/e2011-20015-6} {\bibfield  {journal} {\bibinfo  {journal} {Eur.
  Phys. J. D}\ }\textbf {\bibinfo {volume} {65}},\ \bibinfo {pages} {3}
  (\bibinfo {year} {2011})}\BibitemShut {NoStop}%
\bibitem [{\citenamefont {McCarron}\ \emph {et~al.}(2011)\citenamefont
  {McCarron}, \citenamefont {Cho}, \citenamefont {Jenkin}, \citenamefont
  {K\"oppinger},\ and\ \citenamefont {Cornish}}]{mccarron_11}%
  \BibitemOpen
  \bibfield  {author} {\bibinfo {author} {\bibfnamefont {D.~J.}\ \bibnamefont
  {McCarron}}, \bibinfo {author} {\bibfnamefont {H.~W.}\ \bibnamefont {Cho}},
  \bibinfo {author} {\bibfnamefont {D.~L.}\ \bibnamefont {Jenkin}}, \bibinfo
  {author} {\bibfnamefont {M.~P.}\ \bibnamefont {K\"oppinger}}, \ and\ \bibinfo
  {author} {\bibfnamefont {S.~L.}\ \bibnamefont {Cornish}},\ }\href {\doibase
  10.1103/PhysRevA.84.011603} {\bibfield  {journal} {\bibinfo  {journal} {Phys.
  Rev. A}\ }\textbf {\bibinfo {volume} {84}},\ \bibinfo {pages} {011603(R)}
  (\bibinfo {year} {2011})}\BibitemShut {NoStop}%
\bibitem [{\citenamefont {Pasquiou}\ \emph {et~al.}(2013)\citenamefont
  {Pasquiou}, \citenamefont {Bayerle}, \citenamefont {Tzanova}, \citenamefont
  {Stellmer}, \citenamefont {Szczepkowski}, \citenamefont {Parigger},
  \citenamefont {Grimm},\ and\ \citenamefont {Schreck}}]{pasquiou_13}%
  \BibitemOpen
  \bibfield  {author} {\bibinfo {author} {\bibfnamefont {B.}~\bibnamefont
  {Pasquiou}}, \bibinfo {author} {\bibfnamefont {A.}~\bibnamefont {Bayerle}},
  \bibinfo {author} {\bibfnamefont {S.~M.}\ \bibnamefont {Tzanova}}, \bibinfo
  {author} {\bibfnamefont {S.}~\bibnamefont {Stellmer}}, \bibinfo {author}
  {\bibfnamefont {J.}~\bibnamefont {Szczepkowski}}, \bibinfo {author}
  {\bibfnamefont {M.}~\bibnamefont {Parigger}}, \bibinfo {author}
  {\bibfnamefont {R.}~\bibnamefont {Grimm}}, \ and\ \bibinfo {author}
  {\bibfnamefont {F.}~\bibnamefont {Schreck}},\ }\href {\doibase
  10.1103/PhysRevA.88.023601} {\bibfield  {journal} {\bibinfo  {journal} {Phys.
  Rev. A}\ }\textbf {\bibinfo {volume} {88}},\ \bibinfo {pages} {023601}
  (\bibinfo {year} {2013})}\BibitemShut {NoStop}%
\bibitem [{\citenamefont {Papp}\ \emph {et~al.}(2008)\citenamefont {Papp},
  \citenamefont {Pino},\ and\ \citenamefont {Wieman}}]{papp_08}%
  \BibitemOpen
  \bibfield  {author} {\bibinfo {author} {\bibfnamefont {S.~B.}\ \bibnamefont
  {Papp}}, \bibinfo {author} {\bibfnamefont {J.~M.}\ \bibnamefont {Pino}}, \
  and\ \bibinfo {author} {\bibfnamefont {C.~E.}\ \bibnamefont {Wieman}},\
  }\href {\doibase 10.1103/PhysRevLett.101.040402} {\bibfield  {journal}
  {\bibinfo  {journal} {Phys. Rev. Lett.}\ }\textbf {\bibinfo {volume} {101}},\
  \bibinfo {pages} {040402} (\bibinfo {year} {2008})}\BibitemShut {NoStop}%
\bibitem [{\citenamefont {Tojo}\ \emph {et~al.}(2010)\citenamefont {Tojo},
  \citenamefont {Taguchi}, \citenamefont {Masuyama}, \citenamefont {Hayashi},
  \citenamefont {Saito},\ and\ \citenamefont {Hirano}}]{tojo_10}%
  \BibitemOpen
  \bibfield  {author} {\bibinfo {author} {\bibfnamefont {S.}~\bibnamefont
  {Tojo}}, \bibinfo {author} {\bibfnamefont {Y.}~\bibnamefont {Taguchi}},
  \bibinfo {author} {\bibfnamefont {Y.}~\bibnamefont {Masuyama}}, \bibinfo
  {author} {\bibfnamefont {T.}~\bibnamefont {Hayashi}}, \bibinfo {author}
  {\bibfnamefont {H.}~\bibnamefont {Saito}}, \ and\ \bibinfo {author}
  {\bibfnamefont {T.}~\bibnamefont {Hirano}},\ }\href {\doibase
  10.1103/PhysRevA.82.033609} {\bibfield  {journal} {\bibinfo  {journal} {Phys.
  Rev. A}\ }\textbf {\bibinfo {volume} {82}},\ \bibinfo {pages} {033609}
  (\bibinfo {year} {2010})}\BibitemShut {NoStop}%
\bibitem [{\citenamefont {Wilson}\ \emph {et~al.}(2021)\citenamefont {Wilson},
  \citenamefont {Guttridge}, \citenamefont {Segal},\ and\ \citenamefont
  {Cornish}}]{wilson_21}%
  \BibitemOpen
  \bibfield  {author} {\bibinfo {author} {\bibfnamefont {K.~E.}\ \bibnamefont
  {Wilson}}, \bibinfo {author} {\bibfnamefont {A.}~\bibnamefont {Guttridge}},
  \bibinfo {author} {\bibfnamefont {J.}~\bibnamefont {Segal}}, \ and\ \bibinfo
  {author} {\bibfnamefont {S.~L.}\ \bibnamefont {Cornish}},\ }\href {\doibase
  10.1103/PhysRevA.103.033306} {\bibfield  {journal} {\bibinfo  {journal}
  {Phys. Rev. A}\ }\textbf {\bibinfo {volume} {103}},\ \bibinfo {pages}
  {033306} (\bibinfo {year} {2021})}\BibitemShut {NoStop}%
\bibitem [{\citenamefont {Warner}\ \emph {et~al.}(2021)\citenamefont {Warner},
  \citenamefont {Lam}, \citenamefont {Bigagli}, \citenamefont {Liu},
  \citenamefont {Stevenson},\ and\ \citenamefont {Will}}]{warner_21}%
  \BibitemOpen
  \bibfield  {author} {\bibinfo {author} {\bibfnamefont {C.}~\bibnamefont
  {Warner}}, \bibinfo {author} {\bibfnamefont {A.~Z.}\ \bibnamefont {Lam}},
  \bibinfo {author} {\bibfnamefont {N.}~\bibnamefont {Bigagli}}, \bibinfo
  {author} {\bibfnamefont {H.~C.}\ \bibnamefont {Liu}}, \bibinfo {author}
  {\bibfnamefont {I.}~\bibnamefont {Stevenson}}, \ and\ \bibinfo {author}
  {\bibfnamefont {S.}~\bibnamefont {Will}},\ }\href {\doibase
  10.1103/PhysRevA.104.033302} {\bibfield  {journal} {\bibinfo  {journal}
  {Phys. Rev. A}\ }\textbf {\bibinfo {volume} {104}},\ \bibinfo {pages}
  {033302} (\bibinfo {year} {2021})}\BibitemShut {NoStop}%
\bibitem [{\citenamefont {Farolfi}\ \emph
  {et~al.}(2021{\natexlab{a}})\citenamefont {Farolfi}, \citenamefont
  {Zenesini}, \citenamefont {Trypogeorgos}, \citenamefont {Mordini},
  \citenamefont {Gallem\'\i}, \citenamefont {Roy}, \citenamefont {Recati},
  \citenamefont {Lamporesi},\ and\ \citenamefont {Ferrari}}]{farolfi-2020}%
  \BibitemOpen
  \bibfield  {author} {\bibinfo {author} {\bibfnamefont {A.}~\bibnamefont
  {Farolfi}}, \bibinfo {author} {\bibfnamefont {A.}~\bibnamefont {Zenesini}},
  \bibinfo {author} {\bibfnamefont {D.}~\bibnamefont {Trypogeorgos}}, \bibinfo
  {author} {\bibfnamefont {C.}~\bibnamefont {Mordini}}, \bibinfo {author}
  {\bibfnamefont {A.}~\bibnamefont {Gallem\'\i}}, \bibinfo {author}
  {\bibfnamefont {A.}~\bibnamefont {Roy}}, \bibinfo {author} {\bibfnamefont
  {A.}~\bibnamefont {Recati}}, \bibinfo {author} {\bibfnamefont
  {G.}~\bibnamefont {Lamporesi}}, \ and\ \bibinfo {author} {\bibfnamefont
  {G.}~\bibnamefont {Ferrari}},\ }\href {\doibase 10.1038/s41567-021-01369-y}
  {\bibfield  {journal} {\bibinfo  {journal} {Nat. Physics}\ }\textbf {\bibinfo
  {volume} {17}},\ \bibinfo {pages} {1359} (\bibinfo {year}
  {2021}{\natexlab{a}})}\BibitemShut {NoStop}%
\bibitem [{\citenamefont {Sachdev}(2011)}]{sachdev_2011}%
  \BibitemOpen
  \bibfield  {author} {\bibinfo {author} {\bibfnamefont {S.}~\bibnamefont
  {Sachdev}},\ }\href {\doibase 10.1017/CBO9780511973765} {\emph {\bibinfo
  {title} {Quantum Phase Transitions}}},\ \bibinfo {edition} {2nd}\ ed.\
  (\bibinfo  {publisher} {Cambridge University Press},\ \bibinfo {year}
  {2011})\BibitemShut {NoStop}%
\bibitem [{\citenamefont {Zurek}\ \emph {et~al.}(2005)\citenamefont {Zurek},
  \citenamefont {Dorner},\ and\ \citenamefont {Zoller}}]{zurek_05}%
  \BibitemOpen
  \bibfield  {author} {\bibinfo {author} {\bibfnamefont {W.~H.}\ \bibnamefont
  {Zurek}}, \bibinfo {author} {\bibfnamefont {U.}~\bibnamefont {Dorner}}, \
  and\ \bibinfo {author} {\bibfnamefont {P.}~\bibnamefont {Zoller}},\ }\href
  {\doibase 10.1103/PhysRevLett.95.105701} {\bibfield  {journal} {\bibinfo
  {journal} {Phys. Rev. Lett.}\ }\textbf {\bibinfo {volume} {95}},\ \bibinfo
  {pages} {105701} (\bibinfo {year} {2005})}\BibitemShut {NoStop}%
\bibitem [{\citenamefont {Abad}\ and\ \citenamefont {Recati}(2013)}]{abad_13}%
  \BibitemOpen
  \bibfield  {author} {\bibinfo {author} {\bibfnamefont {M.}~\bibnamefont
  {Abad}}\ and\ \bibinfo {author} {\bibfnamefont {A.}~\bibnamefont {Recati}},\
  }\href {\doibase 10.1140/epjd/e2013-40053-2} {\bibfield  {journal} {\bibinfo
  {journal} {Eur. Phys. J. D}\ }\textbf {\bibinfo {volume} {67}},\ \bibinfo
  {pages} {148} (\bibinfo {year} {2013})}\BibitemShut {NoStop}%
\bibitem [{\citenamefont {Kawaguchi}\ and\ \citenamefont
  {Ueda}(2012{\natexlab{b}})}]{kawaguchi_12}%
  \BibitemOpen
  \bibfield  {author} {\bibinfo {author} {\bibfnamefont {Y.}~\bibnamefont
  {Kawaguchi}}\ and\ \bibinfo {author} {\bibfnamefont {M.}~\bibnamefont
  {Ueda}},\ }\href {\doibase https://doi.org/10.1016/j.physrep.2012.07.005}
  {\bibfield  {journal} {\bibinfo  {journal} {Phys. Rep.}\ }\textbf {\bibinfo
  {volume} {520}},\ \bibinfo {pages} {253} (\bibinfo {year}
  {2012}{\natexlab{b}})}\BibitemShut {NoStop}%
\bibitem [{\citenamefont {Hall}\ \emph {et~al.}(1998)\citenamefont {Hall},
  \citenamefont {Matthews}, \citenamefont {Ensher}, \citenamefont {Wieman},\
  and\ \citenamefont {Cornell}}]{hall_98}%
  \BibitemOpen
  \bibfield  {author} {\bibinfo {author} {\bibfnamefont {D.~S.}\ \bibnamefont
  {Hall}}, \bibinfo {author} {\bibfnamefont {M.~R.}\ \bibnamefont {Matthews}},
  \bibinfo {author} {\bibfnamefont {J.~R.}\ \bibnamefont {Ensher}}, \bibinfo
  {author} {\bibfnamefont {C.~E.}\ \bibnamefont {Wieman}}, \ and\ \bibinfo
  {author} {\bibfnamefont {E.~A.}\ \bibnamefont {Cornell}},\ }\href {\doibase
  10.1103/PhysRevLett.81.1539} {\bibfield  {journal} {\bibinfo  {journal}
  {Phys. Rev. Lett.}\ }\textbf {\bibinfo {volume} {81}},\ \bibinfo {pages}
  {1539} (\bibinfo {year} {1998})}\BibitemShut {NoStop}%
\bibitem [{\citenamefont {Myatt}\ \emph {et~al.}(1997)\citenamefont {Myatt},
  \citenamefont {Burt}, \citenamefont {Ghrist}, \citenamefont {Cornell},\ and\
  \citenamefont {Wieman}}]{myatt_97}%
  \BibitemOpen
  \bibfield  {author} {\bibinfo {author} {\bibfnamefont {C.~J.}\ \bibnamefont
  {Myatt}}, \bibinfo {author} {\bibfnamefont {E.~A.}\ \bibnamefont {Burt}},
  \bibinfo {author} {\bibfnamefont {R.~W.}\ \bibnamefont {Ghrist}}, \bibinfo
  {author} {\bibfnamefont {E.~A.}\ \bibnamefont {Cornell}}, \ and\ \bibinfo
  {author} {\bibfnamefont {C.~E.}\ \bibnamefont {Wieman}},\ }\href {\doibase
  10.1103/PhysRevLett.78.586} {\bibfield  {journal} {\bibinfo  {journal} {Phys.
  Rev. Lett.}\ }\textbf {\bibinfo {volume} {78}},\ \bibinfo {pages} {586}
  (\bibinfo {year} {1997})}\BibitemShut {NoStop}%
\bibitem [{\citenamefont {Barrett}\ \emph {et~al.}(2001)\citenamefont
  {Barrett}, \citenamefont {Sauer},\ and\ \citenamefont
  {Chapman}}]{barrett_01}%
  \BibitemOpen
  \bibfield  {author} {\bibinfo {author} {\bibfnamefont {M.~D.}\ \bibnamefont
  {Barrett}}, \bibinfo {author} {\bibfnamefont {J.~A.}\ \bibnamefont {Sauer}},
  \ and\ \bibinfo {author} {\bibfnamefont {M.~S.}\ \bibnamefont {Chapman}},\
  }\href {\doibase 10.1103/PhysRevLett.87.010404} {\bibfield  {journal}
  {\bibinfo  {journal} {Phys. Rev. Lett.}\ }\textbf {\bibinfo {volume} {87}},\
  \bibinfo {pages} {010404} (\bibinfo {year} {2001})}\BibitemShut {NoStop}%
\bibitem [{\citenamefont {Semeghini}\ \emph {et~al.}(2018)\citenamefont
  {Semeghini}, \citenamefont {Ferioli}, \citenamefont {Masi}, \citenamefont
  {Mazzinghi}, \citenamefont {Wolswijk}, \citenamefont {Minardi}, \citenamefont
  {Modugno}, \citenamefont {Modugno}, \citenamefont {Inguscio},\ and\
  \citenamefont {Fattori}}]{semeghini_18}%
  \BibitemOpen
  \bibfield  {author} {\bibinfo {author} {\bibfnamefont {G.}~\bibnamefont
  {Semeghini}}, \bibinfo {author} {\bibfnamefont {G.}~\bibnamefont {Ferioli}},
  \bibinfo {author} {\bibfnamefont {L.}~\bibnamefont {Masi}}, \bibinfo {author}
  {\bibfnamefont {C.}~\bibnamefont {Mazzinghi}}, \bibinfo {author}
  {\bibfnamefont {L.}~\bibnamefont {Wolswijk}}, \bibinfo {author}
  {\bibfnamefont {F.}~\bibnamefont {Minardi}}, \bibinfo {author} {\bibfnamefont
  {M.}~\bibnamefont {Modugno}}, \bibinfo {author} {\bibfnamefont
  {G.}~\bibnamefont {Modugno}}, \bibinfo {author} {\bibfnamefont
  {M.}~\bibnamefont {Inguscio}}, \ and\ \bibinfo {author} {\bibfnamefont
  {M.}~\bibnamefont {Fattori}},\ }\href {\doibase
  10.1103/PhysRevLett.120.235301} {\bibfield  {journal} {\bibinfo  {journal}
  {Phys. Rev. Lett.}\ }\textbf {\bibinfo {volume} {120}},\ \bibinfo {pages}
  {235301} (\bibinfo {year} {2018})}\BibitemShut {NoStop}%
\bibitem [{\citenamefont {Stenger}\ \emph {et~al.}(1998)\citenamefont
  {Stenger}, \citenamefont {Inouye}, \citenamefont {Stamper-Kurn},
  \citenamefont {Miesner}, \citenamefont {Chikkatur},\ and\ \citenamefont
  {Ketterle}}]{stenger_98}%
  \BibitemOpen
  \bibfield  {author} {\bibinfo {author} {\bibfnamefont {J.}~\bibnamefont
  {Stenger}}, \bibinfo {author} {\bibfnamefont {S.}~\bibnamefont {Inouye}},
  \bibinfo {author} {\bibfnamefont {D.~M.}\ \bibnamefont {Stamper-Kurn}},
  \bibinfo {author} {\bibfnamefont {H.-J.}\ \bibnamefont {Miesner}}, \bibinfo
  {author} {\bibfnamefont {A.~P.}\ \bibnamefont {Chikkatur}}, \ and\ \bibinfo
  {author} {\bibfnamefont {W.}~\bibnamefont {Ketterle}},\ }\href {\doibase
  10.1038/24567} {\bibfield  {journal} {\bibinfo  {journal} {Nature (London)}\
  }\textbf {\bibinfo {volume} {396}},\ \bibinfo {pages} {345} (\bibinfo {year}
  {1998})}\BibitemShut {NoStop}%
\bibitem [{\citenamefont {Kang}\ \emph {et~al.}(2019)\citenamefont {Kang},
  \citenamefont {Seo}, \citenamefont {Takeuchi},\ and\ \citenamefont
  {Shin}}]{kang_19}%
  \BibitemOpen
  \bibfield  {author} {\bibinfo {author} {\bibfnamefont {S.}~\bibnamefont
  {Kang}}, \bibinfo {author} {\bibfnamefont {S.~W.}\ \bibnamefont {Seo}},
  \bibinfo {author} {\bibfnamefont {H.}~\bibnamefont {Takeuchi}}, \ and\
  \bibinfo {author} {\bibfnamefont {Y.}~\bibnamefont {Shin}},\ }\href {\doibase
  10.1103/PhysRevLett.122.095301} {\bibfield  {journal} {\bibinfo  {journal}
  {Phys. Rev. Lett.}\ }\textbf {\bibinfo {volume} {122}},\ \bibinfo {pages}
  {095301} (\bibinfo {year} {2019})}\BibitemShut {NoStop}%
\bibitem [{\citenamefont {Bookjans}\ \emph {et~al.}(2011)\citenamefont
  {Bookjans}, \citenamefont {Vinit},\ and\ \citenamefont
  {Raman}}]{bookjans_11}%
  \BibitemOpen
  \bibfield  {author} {\bibinfo {author} {\bibfnamefont {E.~M.}\ \bibnamefont
  {Bookjans}}, \bibinfo {author} {\bibfnamefont {A.}~\bibnamefont {Vinit}}, \
  and\ \bibinfo {author} {\bibfnamefont {C.}~\bibnamefont {Raman}},\ }\href
  {\doibase 10.1103/PhysRevLett.107.195306} {\bibfield  {journal} {\bibinfo
  {journal} {Phys. Rev. Lett.}\ }\textbf {\bibinfo {volume} {107}},\ \bibinfo
  {pages} {195306} (\bibinfo {year} {2011})}\BibitemShut {NoStop}%
\bibitem [{\citenamefont {Luo}\ \emph {et~al.}(2017)\citenamefont {Luo},
  \citenamefont {Zou}, \citenamefont {Wu}, \citenamefont {Liu}, \citenamefont
  {Han}, \citenamefont {Tey},\ and\ \citenamefont {You}}]{luo_17}%
  \BibitemOpen
  \bibfield  {author} {\bibinfo {author} {\bibfnamefont {X.-Y.}\ \bibnamefont
  {Luo}}, \bibinfo {author} {\bibfnamefont {Y.-Q.}\ \bibnamefont {Zou}},
  \bibinfo {author} {\bibfnamefont {L.-N.}\ \bibnamefont {Wu}}, \bibinfo
  {author} {\bibfnamefont {Q.}~\bibnamefont {Liu}}, \bibinfo {author}
  {\bibfnamefont {M.-F.}\ \bibnamefont {Han}}, \bibinfo {author} {\bibfnamefont
  {M.~K.}\ \bibnamefont {Tey}}, \ and\ \bibinfo {author} {\bibfnamefont
  {L.}~\bibnamefont {You}},\ }\href {\doibase 10.1126/science.aag1106}
  {\bibfield  {journal} {\bibinfo  {journal} {Science}\ }\textbf {\bibinfo
  {volume} {355}},\ \bibinfo {pages} {620} (\bibinfo {year}
  {2017})}\BibitemShut {NoStop}%
\bibitem [{\citenamefont {Chang}\ \emph {et~al.}(2004)\citenamefont {Chang},
  \citenamefont {Hamley}, \citenamefont {Barrett}, \citenamefont {Sauer},
  \citenamefont {Fortier}, \citenamefont {Zhang}, \citenamefont {You},\ and\
  \citenamefont {Chapman}}]{chang_04}%
  \BibitemOpen
  \bibfield  {author} {\bibinfo {author} {\bibfnamefont {M.-S.}\ \bibnamefont
  {Chang}}, \bibinfo {author} {\bibfnamefont {C.~D.}\ \bibnamefont {Hamley}},
  \bibinfo {author} {\bibfnamefont {M.~D.}\ \bibnamefont {Barrett}}, \bibinfo
  {author} {\bibfnamefont {J.~A.}\ \bibnamefont {Sauer}}, \bibinfo {author}
  {\bibfnamefont {K.~M.}\ \bibnamefont {Fortier}}, \bibinfo {author}
  {\bibfnamefont {W.}~\bibnamefont {Zhang}}, \bibinfo {author} {\bibfnamefont
  {L.}~\bibnamefont {You}}, \ and\ \bibinfo {author} {\bibfnamefont {M.~S.}\
  \bibnamefont {Chapman}},\ }\href {\doibase 10.1103/PhysRevLett.92.140403}
  {\bibfield  {journal} {\bibinfo  {journal} {Phys. Rev. Lett.}\ }\textbf
  {\bibinfo {volume} {92}},\ \bibinfo {pages} {140403} (\bibinfo {year}
  {2004})}\BibitemShut {NoStop}%
\bibitem [{\citenamefont {Schmaljohann}\ \emph {et~al.}(2004)\citenamefont
  {Schmaljohann}, \citenamefont {Erhard}, \citenamefont {Kronj{\"a}ger},
  \citenamefont {Sengstock},\ and\ \citenamefont {Bongs}}]{schmaljohann_04}%
  \BibitemOpen
  \bibfield  {author} {\bibinfo {author} {\bibfnamefont {H.}~\bibnamefont
  {Schmaljohann}}, \bibinfo {author} {\bibfnamefont {M.}~\bibnamefont
  {Erhard}}, \bibinfo {author} {\bibfnamefont {J.}~\bibnamefont
  {Kronj{\"a}ger}}, \bibinfo {author} {\bibfnamefont {K.}~\bibnamefont
  {Sengstock}}, \ and\ \bibinfo {author} {\bibfnamefont {K.}~\bibnamefont
  {Bongs}},\ }\href {\doibase 10.1007/s00340-004-1664-6} {\bibfield  {journal}
  {\bibinfo  {journal} {Applied Physics B}\ }\textbf {\bibinfo {volume} {79}},\
  \bibinfo {pages} {1001} (\bibinfo {year} {2004})}\BibitemShut {NoStop}%
\bibitem [{\citenamefont {Zhai}(2012)}]{zhai_12}%
  \BibitemOpen
  \bibfield  {author} {\bibinfo {author} {\bibfnamefont {H.}~\bibnamefont
  {Zhai}},\ }\href {\doibase 10.1142/S0217979212300010} {\bibfield  {journal}
  {\bibinfo  {journal} {International Journal of Modern Physics B}\ }\textbf
  {\bibinfo {volume} {26}},\ \bibinfo {pages} {1230001} (\bibinfo {year}
  {2012})}\BibitemShut {NoStop}%
\bibitem [{\citenamefont {Zhai}(2015)}]{zhai_2015}%
  \BibitemOpen
  \bibfield  {author} {\bibinfo {author} {\bibfnamefont {H.}~\bibnamefont
  {Zhai}},\ }\href {\doibase 10.1088/0034-4885/78/2/026001} {\bibfield
  {journal} {\bibinfo  {journal} {Reports on Progress in Physics}\ }\textbf
  {\bibinfo {volume} {78}},\ \bibinfo {pages} {026001} (\bibinfo {year}
  {2015})}\BibitemShut {NoStop}%
\bibitem [{\citenamefont {Lin}\ \emph {et~al.}(2011)\citenamefont {Lin},
  \citenamefont {Jim{\'e}nez-Garc{\'i}a},\ and\ \citenamefont
  {Spielman}}]{lin_11}%
  \BibitemOpen
  \bibfield  {author} {\bibinfo {author} {\bibfnamefont {Y.-J.}\ \bibnamefont
  {Lin}}, \bibinfo {author} {\bibfnamefont {K.}~\bibnamefont
  {Jim{\'e}nez-Garc{\'i}a}}, \ and\ \bibinfo {author} {\bibfnamefont {I.~B.}\
  \bibnamefont {Spielman}},\ }\href {\doibase 10.1038/nature09887} {\bibfield
  {journal} {\bibinfo  {journal} {Nature}\ }\textbf {\bibinfo {volume} {471}},\
  \bibinfo {pages} {83} (\bibinfo {year} {2011})}\BibitemShut {NoStop}%
\bibitem [{\citenamefont {Kawaguchi}\ \emph {et~al.}(2010)\citenamefont
  {Kawaguchi}, \citenamefont {Kobayashi}, \citenamefont {Nitta},\ and\
  \citenamefont {Ueda}}]{kawaguchi_10}%
  \BibitemOpen
  \bibfield  {author} {\bibinfo {author} {\bibfnamefont {Y.}~\bibnamefont
  {Kawaguchi}}, \bibinfo {author} {\bibfnamefont {M.}~\bibnamefont
  {Kobayashi}}, \bibinfo {author} {\bibfnamefont {M.}~\bibnamefont {Nitta}}, \
  and\ \bibinfo {author} {\bibfnamefont {M.}~\bibnamefont {Ueda}},\ }\href
  {\doibase 10.1143/PTPS.186.455} {\bibfield  {journal} {\bibinfo  {journal}
  {Progress of Theoretical Physics Supplement}\ }\textbf {\bibinfo {volume}
  {186}},\ \bibinfo {pages} {455} (\bibinfo {year} {2010})}\BibitemShut
  {NoStop}%
\bibitem [{\citenamefont {Zibold}\ \emph {et~al.}(2010)\citenamefont {Zibold},
  \citenamefont {Nicklas}, \citenamefont {Gross},\ and\ \citenamefont
  {Oberthaler}}]{zibold_10}%
  \BibitemOpen
  \bibfield  {author} {\bibinfo {author} {\bibfnamefont {T.}~\bibnamefont
  {Zibold}}, \bibinfo {author} {\bibfnamefont {E.}~\bibnamefont {Nicklas}},
  \bibinfo {author} {\bibfnamefont {C.}~\bibnamefont {Gross}}, \ and\ \bibinfo
  {author} {\bibfnamefont {M.~K.}\ \bibnamefont {Oberthaler}},\ }\href
  {\doibase 10.1103/PhysRevLett.105.204101} {\bibfield  {journal} {\bibinfo
  {journal} {Phys. Rev. Lett.}\ }\textbf {\bibinfo {volume} {105}},\ \bibinfo
  {pages} {204101} (\bibinfo {year} {2010})}\BibitemShut {NoStop}%
\bibitem [{\citenamefont {Nicklas}\ \emph {et~al.}(2015)\citenamefont
  {Nicklas}, \citenamefont {Karl}, \citenamefont {H\"ofer}, \citenamefont
  {Johnson}, \citenamefont {Muessel}, \citenamefont {Strobel}, \citenamefont
  {Tomkovi\ifmmode~\check{c}\else \v{c}\fi{}}, \citenamefont {Gasenzer},\ and\
  \citenamefont {Oberthaler}}]{nicklas_15}%
  \BibitemOpen
  \bibfield  {author} {\bibinfo {author} {\bibfnamefont {E.}~\bibnamefont
  {Nicklas}}, \bibinfo {author} {\bibfnamefont {M.}~\bibnamefont {Karl}},
  \bibinfo {author} {\bibfnamefont {M.}~\bibnamefont {H\"ofer}}, \bibinfo
  {author} {\bibfnamefont {A.}~\bibnamefont {Johnson}}, \bibinfo {author}
  {\bibfnamefont {W.}~\bibnamefont {Muessel}}, \bibinfo {author} {\bibfnamefont
  {H.}~\bibnamefont {Strobel}}, \bibinfo {author} {\bibfnamefont
  {J.}~\bibnamefont {Tomkovi\ifmmode~\check{c}\else \v{c}\fi{}}}, \bibinfo
  {author} {\bibfnamefont {T.}~\bibnamefont {Gasenzer}}, \ and\ \bibinfo
  {author} {\bibfnamefont {M.~K.}\ \bibnamefont {Oberthaler}},\ }\href
  {\doibase 10.1103/PhysRevLett.115.245301} {\bibfield  {journal} {\bibinfo
  {journal} {Phys. Rev. Lett.}\ }\textbf {\bibinfo {volume} {115}},\ \bibinfo
  {pages} {245301} (\bibinfo {year} {2015})}\BibitemShut {NoStop}%
\bibitem [{\citenamefont {Kasamatsu}\ \emph {et~al.}(2005)\citenamefont
  {Kasamatsu}, \citenamefont {Tsubota},\ and\ \citenamefont
  {Ueda}}]{kasamatsu_05}%
  \BibitemOpen
  \bibfield  {author} {\bibinfo {author} {\bibfnamefont {K.}~\bibnamefont
  {Kasamatsu}}, \bibinfo {author} {\bibfnamefont {M.}~\bibnamefont {Tsubota}},
  \ and\ \bibinfo {author} {\bibfnamefont {M.}~\bibnamefont {Ueda}},\ }\href
  {\doibase 10.1142/S0217979205029602} {\bibfield  {journal} {\bibinfo
  {journal} {International Journal of Modern Physics B}\ }\textbf {\bibinfo
  {volume} {19}},\ \bibinfo {pages} {1835} (\bibinfo {year}
  {2005})}\BibitemShut {NoStop}%
\bibitem [{\citenamefont {Nicklas}\ \emph {et~al.}(2011)\citenamefont
  {Nicklas}, \citenamefont {Strobel}, \citenamefont {Zibold}, \citenamefont
  {Gross}, \citenamefont {Malomed}, \citenamefont {Kevrekidis},\ and\
  \citenamefont {Oberthaler}}]{nicklas_11}%
  \BibitemOpen
  \bibfield  {author} {\bibinfo {author} {\bibfnamefont {E.}~\bibnamefont
  {Nicklas}}, \bibinfo {author} {\bibfnamefont {H.}~\bibnamefont {Strobel}},
  \bibinfo {author} {\bibfnamefont {T.}~\bibnamefont {Zibold}}, \bibinfo
  {author} {\bibfnamefont {C.}~\bibnamefont {Gross}}, \bibinfo {author}
  {\bibfnamefont {B.~A.}\ \bibnamefont {Malomed}}, \bibinfo {author}
  {\bibfnamefont {P.~G.}\ \bibnamefont {Kevrekidis}}, \ and\ \bibinfo {author}
  {\bibfnamefont {M.~K.}\ \bibnamefont {Oberthaler}},\ }\href {\doibase
  10.1103/PhysRevLett.107.193001} {\bibfield  {journal} {\bibinfo  {journal}
  {Phys. Rev. Lett.}\ }\textbf {\bibinfo {volume} {107}},\ \bibinfo {pages}
  {193001} (\bibinfo {year} {2011})}\BibitemShut {NoStop}%
\bibitem [{\citenamefont {Gross}\ \emph {et~al.}(2011)\citenamefont {Gross},
  \citenamefont {Strobel}, \citenamefont {Nicklas}, \citenamefont {Zibold},
  \citenamefont {Bar-Gill}, \citenamefont {Kurizki},\ and\ \citenamefont
  {Oberthaler}}]{gross_11}%
  \BibitemOpen
  \bibfield  {author} {\bibinfo {author} {\bibfnamefont {C.}~\bibnamefont
  {Gross}}, \bibinfo {author} {\bibfnamefont {H.}~\bibnamefont {Strobel}},
  \bibinfo {author} {\bibfnamefont {E.}~\bibnamefont {Nicklas}}, \bibinfo
  {author} {\bibfnamefont {T.}~\bibnamefont {Zibold}}, \bibinfo {author}
  {\bibfnamefont {N.}~\bibnamefont {Bar-Gill}}, \bibinfo {author}
  {\bibfnamefont {G.}~\bibnamefont {Kurizki}}, \ and\ \bibinfo {author}
  {\bibfnamefont {M.~K.}\ \bibnamefont {Oberthaler}},\ }\href {\doibase
  10.1038/nature10654} {\bibfield  {journal} {\bibinfo  {journal} {Nature}\
  }\textbf {\bibinfo {volume} {480}},\ \bibinfo {pages} {219} (\bibinfo {year}
  {2011})}\BibitemShut {NoStop}%
\bibitem [{\citenamefont {L{\"u}cke}\ \emph {et~al.}(2011)\citenamefont
  {L{\"u}cke}, \citenamefont {Scherer}, \citenamefont {Kruse}, \citenamefont
  {Pezzé}, \citenamefont {Deuretzbacher}, \citenamefont {Hyllus},
  \citenamefont {Topic}, \citenamefont {Peise}, \citenamefont {Ertmer},
  \citenamefont {Arlt}, \citenamefont {Santos}, \citenamefont {Smerzi},\ and\
  \citenamefont {Klempt}}]{lucke_11}%
  \BibitemOpen
  \bibfield  {author} {\bibinfo {author} {\bibfnamefont {B.}~\bibnamefont
  {L{\"u}cke}}, \bibinfo {author} {\bibfnamefont {M.}~\bibnamefont {Scherer}},
  \bibinfo {author} {\bibfnamefont {J.}~\bibnamefont {Kruse}}, \bibinfo
  {author} {\bibfnamefont {L.}~\bibnamefont {Pezzé}}, \bibinfo {author}
  {\bibfnamefont {F.}~\bibnamefont {Deuretzbacher}}, \bibinfo {author}
  {\bibfnamefont {P.}~\bibnamefont {Hyllus}}, \bibinfo {author} {\bibfnamefont
  {O.}~\bibnamefont {Topic}}, \bibinfo {author} {\bibfnamefont
  {J.}~\bibnamefont {Peise}}, \bibinfo {author} {\bibfnamefont
  {W.}~\bibnamefont {Ertmer}}, \bibinfo {author} {\bibfnamefont
  {J.}~\bibnamefont {Arlt}}, \bibinfo {author} {\bibfnamefont {L.}~\bibnamefont
  {Santos}}, \bibinfo {author} {\bibfnamefont {A.}~\bibnamefont {Smerzi}}, \
  and\ \bibinfo {author} {\bibfnamefont {C.}~\bibnamefont {Klempt}},\ }\href
  {\doibase 10.1126/science.1208798} {\bibfield  {journal} {\bibinfo  {journal}
  {Science}\ }\textbf {\bibinfo {volume} {334}},\ \bibinfo {pages} {773}
  (\bibinfo {year} {2011})}\BibitemShut {NoStop}%
\bibitem [{\citenamefont {Yoshino}\ \emph {et~al.}(2021)\citenamefont
  {Yoshino}, \citenamefont {Furukawa},\ and\ \citenamefont {Ueda}}]{takumi_21}%
  \BibitemOpen
  \bibfield  {author} {\bibinfo {author} {\bibfnamefont {T.}~\bibnamefont
  {Yoshino}}, \bibinfo {author} {\bibfnamefont {S.}~\bibnamefont {Furukawa}}, \
  and\ \bibinfo {author} {\bibfnamefont {M.}~\bibnamefont {Ueda}},\ }\href
  {\doibase 10.1103/PhysRevA.103.043321} {\bibfield  {journal} {\bibinfo
  {journal} {Phys. Rev. A}\ }\textbf {\bibinfo {volume} {103}},\ \bibinfo
  {pages} {043321} (\bibinfo {year} {2021})}\BibitemShut {NoStop}%
\bibitem [{\citenamefont {Fischer}\ and\ \citenamefont
  {Sch\"utzhold}(2004)}]{fischer_04}%
  \BibitemOpen
  \bibfield  {author} {\bibinfo {author} {\bibfnamefont {U.~R.}\ \bibnamefont
  {Fischer}}\ and\ \bibinfo {author} {\bibfnamefont {R.}~\bibnamefont
  {Sch\"utzhold}},\ }\href {\doibase 10.1103/PhysRevA.70.063615} {\bibfield
  {journal} {\bibinfo  {journal} {Phys. Rev. A}\ }\textbf {\bibinfo {volume}
  {70}},\ \bibinfo {pages} {063615} (\bibinfo {year} {2004})}\BibitemShut
  {NoStop}%
\bibitem [{\citenamefont {Garay}\ \emph {et~al.}(2000)\citenamefont {Garay},
  \citenamefont {Anglin}, \citenamefont {Cirac},\ and\ \citenamefont
  {Zoller}}]{garay_2000}%
  \BibitemOpen
  \bibfield  {author} {\bibinfo {author} {\bibfnamefont {L.~J.}\ \bibnamefont
  {Garay}}, \bibinfo {author} {\bibfnamefont {J.~R.}\ \bibnamefont {Anglin}},
  \bibinfo {author} {\bibfnamefont {J.~I.}\ \bibnamefont {Cirac}}, \ and\
  \bibinfo {author} {\bibfnamefont {P.}~\bibnamefont {Zoller}},\ }\href
  {\doibase 10.1103/PhysRevLett.85.4643} {\bibfield  {journal} {\bibinfo
  {journal} {Phys. Rev. Lett.}\ }\textbf {\bibinfo {volume} {85}},\ \bibinfo
  {pages} {4643} (\bibinfo {year} {2000})}\BibitemShut {NoStop}%
\bibitem [{\citenamefont {Hazzard}\ and\ \citenamefont
  {Mueller}(2011)}]{hazzard_11}%
  \BibitemOpen
  \bibfield  {author} {\bibinfo {author} {\bibfnamefont {K.~R.~A.}\
  \bibnamefont {Hazzard}}\ and\ \bibinfo {author} {\bibfnamefont {E.~J.}\
  \bibnamefont {Mueller}},\ }\href {\doibase 10.1103/PhysRevA.84.013604}
  {\bibfield  {journal} {\bibinfo  {journal} {Phys. Rev. A}\ }\textbf {\bibinfo
  {volume} {84}},\ \bibinfo {pages} {013604} (\bibinfo {year}
  {2011})}\BibitemShut {NoStop}%
\bibitem [{\citenamefont {Christensen}\ \emph {et~al.}(2021)\citenamefont
  {Christensen}, \citenamefont {Vibel}, \citenamefont {Hilliard}, \citenamefont
  {Kruk}, \citenamefont {Pawlowski}, \citenamefont {Hryniuk}, \citenamefont
  {Rzazewski}, \citenamefont {Kristensen},\ and\ \citenamefont
  {Arlt}}]{christensen_21}%
  \BibitemOpen
  \bibfield  {author} {\bibinfo {author} {\bibfnamefont {M.~B.}\ \bibnamefont
  {Christensen}}, \bibinfo {author} {\bibfnamefont {T.}~\bibnamefont {Vibel}},
  \bibinfo {author} {\bibfnamefont {A.~J.}\ \bibnamefont {Hilliard}}, \bibinfo
  {author} {\bibfnamefont {M.~B.}\ \bibnamefont {Kruk}}, \bibinfo {author}
  {\bibfnamefont {K.}~\bibnamefont {Pawlowski}}, \bibinfo {author}
  {\bibfnamefont {D.}~\bibnamefont {Hryniuk}}, \bibinfo {author} {\bibfnamefont
  {K.}~\bibnamefont {Rzazewski}}, \bibinfo {author} {\bibfnamefont {M.~A.}\
  \bibnamefont {Kristensen}}, \ and\ \bibinfo {author} {\bibfnamefont {J.~J.}\
  \bibnamefont {Arlt}},\ }\href {\doibase 10.1103/PhysRevLett.126.153601}
  {\bibfield  {journal} {\bibinfo  {journal} {Phys. Rev. Lett.}\ }\textbf
  {\bibinfo {volume} {126}},\ \bibinfo {pages} {153601} (\bibinfo {year}
  {2021})}\BibitemShut {NoStop}%
\bibitem [{\citenamefont {Sondhi}\ \emph {et~al.}(1997)\citenamefont {Sondhi},
  \citenamefont {Girvin}, \citenamefont {Carini},\ and\ \citenamefont
  {Shahar}}]{sondhi_97}%
  \BibitemOpen
  \bibfield  {author} {\bibinfo {author} {\bibfnamefont {S.~L.}\ \bibnamefont
  {Sondhi}}, \bibinfo {author} {\bibfnamefont {S.~M.}\ \bibnamefont {Girvin}},
  \bibinfo {author} {\bibfnamefont {J.~P.}\ \bibnamefont {Carini}}, \ and\
  \bibinfo {author} {\bibfnamefont {D.}~\bibnamefont {Shahar}},\ }\href
  {\doibase 10.1103/RevModPhys.69.315} {\bibfield  {journal} {\bibinfo
  {journal} {Rev. Mod. Phys.}\ }\textbf {\bibinfo {volume} {69}},\ \bibinfo
  {pages} {315} (\bibinfo {year} {1997})}\BibitemShut {NoStop}%
\bibitem [{\citenamefont {Dutta}\ \emph {et~al.}(2015)\citenamefont {Dutta},
  \citenamefont {Aeppli}, \citenamefont {Chakrabarti}, \citenamefont
  {Divakaran}, \citenamefont {Rosenbaum},\ and\ \citenamefont
  {Sen}}]{dutta_2015}%
  \BibitemOpen
  \bibfield  {author} {\bibinfo {author} {\bibfnamefont {A.}~\bibnamefont
  {Dutta}}, \bibinfo {author} {\bibfnamefont {G.}~\bibnamefont {Aeppli}},
  \bibinfo {author} {\bibfnamefont {B.~K.}\ \bibnamefont {Chakrabarti}},
  \bibinfo {author} {\bibfnamefont {U.}~\bibnamefont {Divakaran}}, \bibinfo
  {author} {\bibfnamefont {T.~F.}\ \bibnamefont {Rosenbaum}}, \ and\ \bibinfo
  {author} {\bibfnamefont {D.}~\bibnamefont {Sen}},\ }\href {\doibase
  10.1017/CBO9781107706057} {\emph {\bibinfo {title} {Quantum Phase Transitions
  in Transverse Field Spin Models: From Statistical Physics to Quantum
  Information}}}\ (\bibinfo  {publisher} {Cambridge University Press},\
  \bibinfo {year} {2015})\BibitemShut {NoStop}%
\bibitem [{\citenamefont {Carr}(2010)}]{carr_10}%
  \BibitemOpen
  \bibfield  {author} {\bibinfo {author} {\bibfnamefont {L.}~\bibnamefont
  {Carr}},\ }\href@noop {} {\emph {\bibinfo {title} {Understanding Quantum
  Phase Transitions}}}\ (\bibinfo  {publisher} {CRC Press},\ \bibinfo {year}
  {2010})\BibitemShut {NoStop}%
\bibitem [{\citenamefont {Chen}\ \emph {et~al.}(2017)\citenamefont {Chen},
  \citenamefont {Liu},\ and\ \citenamefont {Hu}}]{chen_17}%
  \BibitemOpen
  \bibfield  {author} {\bibinfo {author} {\bibfnamefont {X.-L.}\ \bibnamefont
  {Chen}}, \bibinfo {author} {\bibfnamefont {X.-J.}\ \bibnamefont {Liu}}, \
  and\ \bibinfo {author} {\bibfnamefont {H.}~\bibnamefont {Hu}},\ }\href
  {\doibase 10.1103/PhysRevA.96.013625} {\bibfield  {journal} {\bibinfo
  {journal} {Phys. Rev. A}\ }\textbf {\bibinfo {volume} {96}},\ \bibinfo
  {pages} {013625} (\bibinfo {year} {2017})}\BibitemShut {NoStop}%
\bibitem [{\citenamefont {Chen}\ \emph {et~al.}(2018)\citenamefont {Chen},
  \citenamefont {Wang}, \citenamefont {Li}, \citenamefont {Liu},\ and\
  \citenamefont {Hu}}]{chen_18}%
  \BibitemOpen
  \bibfield  {author} {\bibinfo {author} {\bibfnamefont {X.-L.}\ \bibnamefont
  {Chen}}, \bibinfo {author} {\bibfnamefont {J.}~\bibnamefont {Wang}}, \bibinfo
  {author} {\bibfnamefont {Y.}~\bibnamefont {Li}}, \bibinfo {author}
  {\bibfnamefont {X.-J.}\ \bibnamefont {Liu}}, \ and\ \bibinfo {author}
  {\bibfnamefont {H.}~\bibnamefont {Hu}},\ }\href {\doibase
  10.1103/PhysRevA.98.013614} {\bibfield  {journal} {\bibinfo  {journal} {Phys.
  Rev. A}\ }\textbf {\bibinfo {volume} {98}},\ \bibinfo {pages} {013614}
  (\bibinfo {year} {2018})}\BibitemShut {NoStop}%
\bibitem [{\citenamefont {Su}\ \emph {et~al.}(2017)\citenamefont {Su},
  \citenamefont {Liu}, \citenamefont {Gou}, \citenamefont {Liao}, \citenamefont
  {Fialko},\ and\ \citenamefont {Brand}}]{su_17}%
  \BibitemOpen
  \bibfield  {author} {\bibinfo {author} {\bibfnamefont {S.-W.}\ \bibnamefont
  {Su}}, \bibinfo {author} {\bibfnamefont {I.-K.}\ \bibnamefont {Liu}},
  \bibinfo {author} {\bibfnamefont {S.-C.}\ \bibnamefont {Gou}}, \bibinfo
  {author} {\bibfnamefont {R.}~\bibnamefont {Liao}}, \bibinfo {author}
  {\bibfnamefont {O.}~\bibnamefont {Fialko}}, \ and\ \bibinfo {author}
  {\bibfnamefont {J.}~\bibnamefont {Brand}},\ }\href {\doibase
  10.1103/PhysRevA.95.053629} {\bibfield  {journal} {\bibinfo  {journal} {Phys.
  Rev. A}\ }\textbf {\bibinfo {volume} {95}},\ \bibinfo {pages} {053629}
  (\bibinfo {year} {2017})}\BibitemShut {NoStop}%
\bibitem [{\citenamefont {Attanasio}\ and\ \citenamefont
  {Drut}(2020)}]{attanasio_20}%
  \BibitemOpen
  \bibfield  {author} {\bibinfo {author} {\bibfnamefont {F.}~\bibnamefont
  {Attanasio}}\ and\ \bibinfo {author} {\bibfnamefont {J.~E.}\ \bibnamefont
  {Drut}},\ }\href {\doibase 10.1103/PhysRevA.101.033617} {\bibfield  {journal}
  {\bibinfo  {journal} {Phys. Rev. A}\ }\textbf {\bibinfo {volume} {101}},\
  \bibinfo {pages} {033617} (\bibinfo {year} {2020})}\BibitemShut {NoStop}%
\bibitem [{\citenamefont {Liu}\ \emph {et~al.}(2012)\citenamefont {Liu},
  \citenamefont {Fan}, \citenamefont {Zhang}, \citenamefont {Wang},\ and\
  \citenamefont {Liu}}]{liu_12}%
  \BibitemOpen
  \bibfield  {author} {\bibinfo {author} {\bibfnamefont {C.-F.}\ \bibnamefont
  {Liu}}, \bibinfo {author} {\bibfnamefont {H.}~\bibnamefont {Fan}}, \bibinfo
  {author} {\bibfnamefont {Y.-C.}\ \bibnamefont {Zhang}}, \bibinfo {author}
  {\bibfnamefont {D.-S.}\ \bibnamefont {Wang}}, \ and\ \bibinfo {author}
  {\bibfnamefont {W.-M.}\ \bibnamefont {Liu}},\ }\href {\doibase
  10.1103/PhysRevA.86.053616} {\bibfield  {journal} {\bibinfo  {journal} {Phys.
  Rev. A}\ }\textbf {\bibinfo {volume} {86}},\ \bibinfo {pages} {053616}
  (\bibinfo {year} {2012})}\BibitemShut {NoStop}%
\bibitem [{\citenamefont {Su}\ \emph {et~al.}(2012)\citenamefont {Su},
  \citenamefont {Liu}, \citenamefont {Tsai}, \citenamefont {Liu},\ and\
  \citenamefont {Gou}}]{su_12}%
  \BibitemOpen
  \bibfield  {author} {\bibinfo {author} {\bibfnamefont {S.-W.}\ \bibnamefont
  {Su}}, \bibinfo {author} {\bibfnamefont {I.-K.}\ \bibnamefont {Liu}},
  \bibinfo {author} {\bibfnamefont {Y.-C.}\ \bibnamefont {Tsai}}, \bibinfo
  {author} {\bibfnamefont {W.~M.}\ \bibnamefont {Liu}}, \ and\ \bibinfo
  {author} {\bibfnamefont {S.-C.}\ \bibnamefont {Gou}},\ }\href {\doibase
  10.1103/PhysRevA.86.023601} {\bibfield  {journal} {\bibinfo  {journal} {Phys.
  Rev. A}\ }\textbf {\bibinfo {volume} {86}},\ \bibinfo {pages} {023601}
  (\bibinfo {year} {2012})}\BibitemShut {NoStop}%
\bibitem [{\citenamefont {Ji}\ \emph {et~al.}(2014)\citenamefont {Ji},
  \citenamefont {Zhang}, \citenamefont {Zhang}, \citenamefont {Du},
  \citenamefont {Zheng}, \citenamefont {Deng}, \citenamefont {Zhai},
  \citenamefont {Chen},\ and\ \citenamefont {Pan}}]{ji_14}%
  \BibitemOpen
  \bibfield  {author} {\bibinfo {author} {\bibfnamefont {S.-C.}\ \bibnamefont
  {Ji}}, \bibinfo {author} {\bibfnamefont {J.-Y.}\ \bibnamefont {Zhang}},
  \bibinfo {author} {\bibfnamefont {L.}~\bibnamefont {Zhang}}, \bibinfo
  {author} {\bibfnamefont {Z.-D.}\ \bibnamefont {Du}}, \bibinfo {author}
  {\bibfnamefont {W.}~\bibnamefont {Zheng}}, \bibinfo {author} {\bibfnamefont
  {Y.-J.}\ \bibnamefont {Deng}}, \bibinfo {author} {\bibfnamefont
  {H.}~\bibnamefont {Zhai}}, \bibinfo {author} {\bibfnamefont {S.}~\bibnamefont
  {Chen}}, \ and\ \bibinfo {author} {\bibfnamefont {J.-W.}\ \bibnamefont
  {Pan}},\ }\href {\doibase 10.1038/nphys2905} {\bibfield  {journal} {\bibinfo
  {journal} {Nature Physics}\ }\textbf {\bibinfo {volume} {10}},\ \bibinfo
  {pages} {314} (\bibinfo {year} {2014})}\BibitemShut {NoStop}%
\bibitem [{\citenamefont {Gerry}\ and\ \citenamefont
  {Knight}(2004)}]{gerry_knight_2004}%
  \BibitemOpen
  \bibfield  {author} {\bibinfo {author} {\bibfnamefont {C.}~\bibnamefont
  {Gerry}}\ and\ \bibinfo {author} {\bibfnamefont {P.}~\bibnamefont {Knight}},\
  }\href {\doibase 10.1017/CBO9780511791239} {\emph {\bibinfo {title}
  {Introductory Quantum Optics}}}\ (\bibinfo  {publisher} {Cambridge University
  Press},\ \bibinfo {year} {2004})\BibitemShut {NoStop}%
\bibitem [{\citenamefont {Leggett}(2001)}]{leggett_01}%
  \BibitemOpen
  \bibfield  {author} {\bibinfo {author} {\bibfnamefont {A.~J.}\ \bibnamefont
  {Leggett}},\ }\href {\doibase 10.1103/RevModPhys.73.307} {\bibfield
  {journal} {\bibinfo  {journal} {Rev. Mod. Phys.}\ }\textbf {\bibinfo {volume}
  {73}},\ \bibinfo {pages} {307} (\bibinfo {year} {2001})}\BibitemShut
  {NoStop}%
\bibitem [{\citenamefont {Farolfi}\ \emph
  {et~al.}(2021{\natexlab{b}})\citenamefont {Farolfi}, \citenamefont
  {Zenesini}, \citenamefont {Cominotti}, \citenamefont {Trypogeorgos},
  \citenamefont {Recati}, \citenamefont {Lamporesi},\ and\ \citenamefont
  {Ferrari}}]{farolfi_21}%
  \BibitemOpen
  \bibfield  {author} {\bibinfo {author} {\bibfnamefont {A.}~\bibnamefont
  {Farolfi}}, \bibinfo {author} {\bibfnamefont {A.}~\bibnamefont {Zenesini}},
  \bibinfo {author} {\bibfnamefont {R.}~\bibnamefont {Cominotti}}, \bibinfo
  {author} {\bibfnamefont {D.}~\bibnamefont {Trypogeorgos}}, \bibinfo {author}
  {\bibfnamefont {A.}~\bibnamefont {Recati}}, \bibinfo {author} {\bibfnamefont
  {G.}~\bibnamefont {Lamporesi}}, \ and\ \bibinfo {author} {\bibfnamefont
  {G.}~\bibnamefont {Ferrari}},\ }\href {\doibase 10.1103/PhysRevA.104.023326}
  {\bibfield  {journal} {\bibinfo  {journal} {Phys. Rev. A}\ }\textbf {\bibinfo
  {volume} {104}},\ \bibinfo {pages} {023326} (\bibinfo {year}
  {2021}{\natexlab{b}})}\BibitemShut {NoStop}%
\bibitem [{\citenamefont {Gaunt}\ \emph {et~al.}(2013)\citenamefont {Gaunt},
  \citenamefont {Schmidutz}, \citenamefont {Gotlibovych}, \citenamefont
  {Smith},\ and\ \citenamefont {Hadzibabic}}]{gaunt_13}%
  \BibitemOpen
  \bibfield  {author} {\bibinfo {author} {\bibfnamefont {A.~L.}\ \bibnamefont
  {Gaunt}}, \bibinfo {author} {\bibfnamefont {T.~F.}\ \bibnamefont
  {Schmidutz}}, \bibinfo {author} {\bibfnamefont {I.}~\bibnamefont
  {Gotlibovych}}, \bibinfo {author} {\bibfnamefont {R.~P.}\ \bibnamefont
  {Smith}}, \ and\ \bibinfo {author} {\bibfnamefont {Z.}~\bibnamefont
  {Hadzibabic}},\ }\href {\doibase 10.1103/PhysRevLett.110.200406} {\bibfield
  {journal} {\bibinfo  {journal} {Phys. Rev. Lett.}\ }\textbf {\bibinfo
  {volume} {110}},\ \bibinfo {pages} {200406} (\bibinfo {year}
  {2013})}\BibitemShut {NoStop}%
\bibitem [{\citenamefont {Chomaz}\ \emph {et~al.}(2015)\citenamefont {Chomaz},
  \citenamefont {Corman}, \citenamefont {Bienaim{\'e}}, \citenamefont
  {Desbuquois}, \citenamefont {Weitenberg}, \citenamefont {Nascimb{\'e}ne},
  \citenamefont {Beugnon},\ and\ \citenamefont {Dalibard}}]{chomaz_15}%
  \BibitemOpen
  \bibfield  {author} {\bibinfo {author} {\bibfnamefont {L.}~\bibnamefont
  {Chomaz}}, \bibinfo {author} {\bibfnamefont {L.}~\bibnamefont {Corman}},
  \bibinfo {author} {\bibfnamefont {T.}~\bibnamefont {Bienaim{\'e}}}, \bibinfo
  {author} {\bibfnamefont {R.}~\bibnamefont {Desbuquois}}, \bibinfo {author}
  {\bibfnamefont {C.}~\bibnamefont {Weitenberg}}, \bibinfo {author}
  {\bibfnamefont {S.}~\bibnamefont {Nascimb{\'e}ne}}, \bibinfo {author}
  {\bibfnamefont {J.}~\bibnamefont {Beugnon}}, \ and\ \bibinfo {author}
  {\bibfnamefont {J.}~\bibnamefont {Dalibard}},\ }\href {\doibase
  10.1038/ncomms7162} {\bibfield  {journal} {\bibinfo  {journal} {Nat. Comm.}\
  }\textbf {\bibinfo {volume} {6}},\ \bibinfo {pages} {6162} (\bibinfo {year}
  {2015})}\BibitemShut {NoStop}%
\bibitem [{\citenamefont {Pitaevskii}\ and\ \citenamefont
  {Stringari}(2016{\natexlab{b}})}]{pitaevskii_16}%
  \BibitemOpen
  \bibfield  {author} {\bibinfo {author} {\bibfnamefont {L.}~\bibnamefont
  {Pitaevskii}}\ and\ \bibinfo {author} {\bibfnamefont {S.}~\bibnamefont
  {Stringari}},\ }\href {https://books.google.es/books?id=yHByCwAAQBAJ} {\emph
  {\bibinfo {title} {{B}ose-{E}instein Condensation and Superfluidity}}},\
  International Series of Monographs on Physics\ (\bibinfo  {publisher} {OUP
  Oxford},\ \bibinfo {year} {2016})\BibitemShut {NoStop}%
\bibitem [{\citenamefont {Bernier}\ \emph {et~al.}(2014)\citenamefont
  {Bernier}, \citenamefont {Dalla~Torre},\ and\ \citenamefont
  {Demler}}]{bernier_14}%
  \BibitemOpen
  \bibfield  {author} {\bibinfo {author} {\bibfnamefont {N.~R.}\ \bibnamefont
  {Bernier}}, \bibinfo {author} {\bibfnamefont {E.~G.}\ \bibnamefont
  {Dalla~Torre}}, \ and\ \bibinfo {author} {\bibfnamefont {E.}~\bibnamefont
  {Demler}},\ }\href {\doibase 10.1103/PhysRevLett.113.065303} {\bibfield
  {journal} {\bibinfo  {journal} {Phys. Rev. Lett.}\ }\textbf {\bibinfo
  {volume} {113}},\ \bibinfo {pages} {065303} (\bibinfo {year}
  {2014})}\BibitemShut {NoStop}%
\bibitem [{\citenamefont {Roy}\ \emph {et~al.}(2021)\citenamefont {Roy},
  \citenamefont {Ota}, \citenamefont {Recati},\ and\ \citenamefont
  {Dalfovo}}]{roy_21}%
  \BibitemOpen
  \bibfield  {author} {\bibinfo {author} {\bibfnamefont {A.}~\bibnamefont
  {Roy}}, \bibinfo {author} {\bibfnamefont {M.}~\bibnamefont {Ota}}, \bibinfo
  {author} {\bibfnamefont {A.}~\bibnamefont {Recati}}, \ and\ \bibinfo {author}
  {\bibfnamefont {F.}~\bibnamefont {Dalfovo}},\ }\href {\doibase
  10.1103/PhysRevResearch.3.013161} {\bibfield  {journal} {\bibinfo  {journal}
  {Phys. Rev. Research}\ }\textbf {\bibinfo {volume} {3}},\ \bibinfo {pages}
  {013161} (\bibinfo {year} {2021})}\BibitemShut {NoStop}%
\bibitem [{\citenamefont {Recati}\ and\ \citenamefont
  {Piazza}(2019)}]{recati_19}%
  \BibitemOpen
  \bibfield  {author} {\bibinfo {author} {\bibfnamefont {A.}~\bibnamefont
  {Recati}}\ and\ \bibinfo {author} {\bibfnamefont {F.}~\bibnamefont
  {Piazza}},\ }\href {\doibase 10.1103/PhysRevB.99.064505} {\bibfield
  {journal} {\bibinfo  {journal} {Phys. Rev. B}\ }\textbf {\bibinfo {volume}
  {99}},\ \bibinfo {pages} {064505} (\bibinfo {year} {2019})}\BibitemShut
  {NoStop}%
\bibitem [{\citenamefont {Recati}\ and\ \citenamefont
  {Stringari}(2022)}]{recati_21}%
  \BibitemOpen
  \bibfield  {author} {\bibinfo {author} {\bibfnamefont {A.}~\bibnamefont
  {Recati}}\ and\ \bibinfo {author} {\bibfnamefont {S.}~\bibnamefont
  {Stringari}},\ }\href {\doibase 10.1146/annurev-conmatphys-031820-121316}
  {\bibfield  {journal} {\bibinfo  {journal} {Annual Review of Condensed Matter
  Physics}\ }\textbf {\bibinfo {volume} {13}},\ \bibinfo {pages} {407}
  (\bibinfo {year} {2022})}\BibitemShut {NoStop}%
\bibitem [{\citenamefont {Trenkwalder}\ \emph {et~al.}(2016)\citenamefont
  {Trenkwalder}, \citenamefont {Spagnolli}, \citenamefont {Semeghini},
  \citenamefont {Coop}, \citenamefont {Landini}, \citenamefont {Castilho},
  \citenamefont {Pezz{\'e}}, \citenamefont {Modugno}, \citenamefont {Inguscio},
  \citenamefont {Smerzi},\ and\ \citenamefont {Fattori}}]{trenkwalder_16}%
  \BibitemOpen
  \bibfield  {author} {\bibinfo {author} {\bibfnamefont {A.}~\bibnamefont
  {Trenkwalder}}, \bibinfo {author} {\bibfnamefont {G.}~\bibnamefont
  {Spagnolli}}, \bibinfo {author} {\bibfnamefont {G.}~\bibnamefont
  {Semeghini}}, \bibinfo {author} {\bibfnamefont {S.}~\bibnamefont {Coop}},
  \bibinfo {author} {\bibfnamefont {M.}~\bibnamefont {Landini}}, \bibinfo
  {author} {\bibfnamefont {P.}~\bibnamefont {Castilho}}, \bibinfo {author}
  {\bibfnamefont {L.}~\bibnamefont {Pezz{\'e}}}, \bibinfo {author}
  {\bibfnamefont {G.}~\bibnamefont {Modugno}}, \bibinfo {author} {\bibfnamefont
  {M.}~\bibnamefont {Inguscio}}, \bibinfo {author} {\bibfnamefont
  {A.}~\bibnamefont {Smerzi}}, \ and\ \bibinfo {author} {\bibfnamefont
  {M.}~\bibnamefont {Fattori}},\ }\href {\doibase 10.1038/nphys3743} {\bibfield
   {journal} {\bibinfo  {journal} {Nature Physics}\ }\textbf {\bibinfo {volume}
  {12}},\ \bibinfo {pages} {826} (\bibinfo {year} {2016})}\BibitemShut
  {NoStop}%
\bibitem [{\citenamefont {Cominotti}\ \emph
  {et~al.}(2022{\natexlab{a}})\citenamefont {Cominotti}, \citenamefont {Berti},
  \citenamefont {Dulin}, \citenamefont {Rogora}, \citenamefont {Lamporesi},
  \citenamefont {Carusotto}, \citenamefont {Recati}, \citenamefont {Zenesini},\
  and\ \citenamefont {Ferrari}}]{cominotti_22}%
  \BibitemOpen
  \bibfield  {author} {\bibinfo {author} {\bibfnamefont {R.}~\bibnamefont
  {Cominotti}}, \bibinfo {author} {\bibfnamefont {A.}~\bibnamefont {Berti}},
  \bibinfo {author} {\bibfnamefont {C.}~\bibnamefont {Dulin}}, \bibinfo
  {author} {\bibfnamefont {C.}~\bibnamefont {Rogora}}, \bibinfo {author}
  {\bibfnamefont {G.}~\bibnamefont {Lamporesi}}, \bibinfo {author}
  {\bibfnamefont {I.}~\bibnamefont {Carusotto}}, \bibinfo {author}
  {\bibfnamefont {A.}~\bibnamefont {Recati}}, \bibinfo {author} {\bibfnamefont
  {A.}~\bibnamefont {Zenesini}}, \ and\ \bibinfo {author} {\bibfnamefont
  {G.}~\bibnamefont {Ferrari}},\ }\href {\doibase 10.48550/ARXIV.2209.13235}
  {\bibfield  {journal} {\bibinfo  {journal} {arXiv:2209.13235}\ } (\bibinfo
  {year} {2022}{\natexlab{a}}),\ 10.48550/ARXIV.2209.13235}\BibitemShut
  {NoStop}%
\bibitem [{\citenamefont {Stoof}\ and\ \citenamefont
  {Bijlsma}(2001)}]{stoof_01}%
  \BibitemOpen
  \bibfield  {author} {\bibinfo {author} {\bibfnamefont {H.~T.~C.}\
  \bibnamefont {Stoof}}\ and\ \bibinfo {author} {\bibfnamefont {M.~J.}\
  \bibnamefont {Bijlsma}},\ }\href {\doibase 10.1023/A:1017519118408}
  {\bibfield  {journal} {\bibinfo  {journal} {J. Low Temp. Phys.}\ }\textbf
  {\bibinfo {volume} {124}},\ \bibinfo {pages} {431} (\bibinfo {year}
  {2001})}\BibitemShut {NoStop}%
\bibitem [{\citenamefont {Proukakis}\ \emph {et~al.}(2006)\citenamefont
  {Proukakis}, \citenamefont {Schmiedmayer},\ and\ \citenamefont
  {Stoof}}]{proukakis_06}%
  \BibitemOpen
  \bibfield  {author} {\bibinfo {author} {\bibfnamefont {N.~P.}\ \bibnamefont
  {Proukakis}}, \bibinfo {author} {\bibfnamefont {J.}~\bibnamefont
  {Schmiedmayer}}, \ and\ \bibinfo {author} {\bibfnamefont {H.~T.~C.}\
  \bibnamefont {Stoof}},\ }\href {\doibase 10.1103/PhysRevA.73.053603}
  {\bibfield  {journal} {\bibinfo  {journal} {Phys. Rev. A}\ }\textbf {\bibinfo
  {volume} {73}},\ \bibinfo {pages} {053603} (\bibinfo {year}
  {2006})}\BibitemShut {NoStop}%
\bibitem [{\citenamefont {Proukakis}\ and\ \citenamefont
  {Jackson}(2008)}]{proukakis_08}%
  \BibitemOpen
  \bibfield  {author} {\bibinfo {author} {\bibfnamefont {N.~P.}\ \bibnamefont
  {Proukakis}}\ and\ \bibinfo {author} {\bibfnamefont {B.}~\bibnamefont
  {Jackson}},\ }\href {http://stacks.iop.org/0953-4075/41/i=20/a=203002}
  {\bibfield  {journal} {\bibinfo  {journal} {J. Phys. B}\ }\textbf {\bibinfo
  {volume} {41}},\ \bibinfo {pages} {203002} (\bibinfo {year}
  {2008})}\BibitemShut {NoStop}%
\bibitem [{\citenamefont {Blakie}\ \emph {et~al.}(2008)\citenamefont {Blakie},
  \citenamefont {Bradley}, \citenamefont {Davis}, \citenamefont {Ballagh},\
  and\ \citenamefont {Gardiner}}]{blakie_08}%
  \BibitemOpen
  \bibfield  {author} {\bibinfo {author} {\bibfnamefont {P.}~\bibnamefont
  {Blakie}}, \bibinfo {author} {\bibfnamefont {A.}~\bibnamefont {Bradley}},
  \bibinfo {author} {\bibfnamefont {M.}~\bibnamefont {Davis}}, \bibinfo
  {author} {\bibfnamefont {R.}~\bibnamefont {Ballagh}}, \ and\ \bibinfo
  {author} {\bibfnamefont {C.}~\bibnamefont {Gardiner}},\ }\href
  {http://www.tandfonline.com/doi/abs/10.1080/00018730802564254} {\bibfield
  {journal} {\bibinfo  {journal} {Adv. Phys.}\ }\textbf {\bibinfo {volume}
  {57}},\ \bibinfo {pages} {363} (\bibinfo {year} {2008})}\BibitemShut
  {NoStop}%
\bibitem [{\citenamefont {Bradley}\ \emph {et~al.}(2008)\citenamefont
  {Bradley}, \citenamefont {Gardiner},\ and\ \citenamefont
  {Davis}}]{bradley_08}%
  \BibitemOpen
  \bibfield  {author} {\bibinfo {author} {\bibfnamefont {A.~S.}\ \bibnamefont
  {Bradley}}, \bibinfo {author} {\bibfnamefont {C.~W.}\ \bibnamefont
  {Gardiner}}, \ and\ \bibinfo {author} {\bibfnamefont {M.~J.}\ \bibnamefont
  {Davis}},\ }\href {\doibase 10.1103/PhysRevA.77.033616} {\bibfield  {journal}
  {\bibinfo  {journal} {Phys. Rev. A}\ }\textbf {\bibinfo {volume} {77}},\
  \bibinfo {pages} {033616} (\bibinfo {year} {2008})}\BibitemShut {NoStop}%
\bibitem [{\citenamefont {Cockburn}\ and\ \citenamefont
  {Proukakis}(2009)}]{cockburn_09}%
  \BibitemOpen
  \bibfield  {author} {\bibinfo {author} {\bibfnamefont {S.~P.}\ \bibnamefont
  {Cockburn}}\ and\ \bibinfo {author} {\bibfnamefont {N.~P.}\ \bibnamefont
  {Proukakis}},\ }\href {\doibase 10.1134/S1054660X09040057} {\bibfield
  {journal} {\bibinfo  {journal} {Las. Phys.}\ }\textbf {\bibinfo {volume}
  {19}},\ \bibinfo {pages} {558} (\bibinfo {year} {2009})}\BibitemShut
  {NoStop}%
\bibitem [{\citenamefont {Su}\ \emph {et~al.}(2011)\citenamefont {Su},
  \citenamefont {Hsueh}, \citenamefont {Liu}, \citenamefont {Horng},
  \citenamefont {Tsai}, \citenamefont {Gou},\ and\ \citenamefont
  {Liu}}]{su_11}%
  \BibitemOpen
  \bibfield  {author} {\bibinfo {author} {\bibfnamefont {S.-W.}\ \bibnamefont
  {Su}}, \bibinfo {author} {\bibfnamefont {C.-H.}\ \bibnamefont {Hsueh}},
  \bibinfo {author} {\bibfnamefont {I.-K.}\ \bibnamefont {Liu}}, \bibinfo
  {author} {\bibfnamefont {T.-L.}\ \bibnamefont {Horng}}, \bibinfo {author}
  {\bibfnamefont {Y.-C.}\ \bibnamefont {Tsai}}, \bibinfo {author}
  {\bibfnamefont {S.-C.}\ \bibnamefont {Gou}}, \ and\ \bibinfo {author}
  {\bibfnamefont {W.~M.}\ \bibnamefont {Liu}},\ }\href {\doibase
  10.1103/PhysRevA.84.023601} {\bibfield  {journal} {\bibinfo  {journal} {Phys.
  Rev. A}\ }\textbf {\bibinfo {volume} {84}},\ \bibinfo {pages} {023601}
  (\bibinfo {year} {2011})}\BibitemShut {NoStop}%
\bibitem [{\citenamefont {Rooney}\ \emph {et~al.}(2013)\citenamefont {Rooney},
  \citenamefont {Neely}, \citenamefont {Anderson},\ and\ \citenamefont
  {Bradley}}]{rooney_13}%
  \BibitemOpen
  \bibfield  {author} {\bibinfo {author} {\bibfnamefont {S.~J.}\ \bibnamefont
  {Rooney}}, \bibinfo {author} {\bibfnamefont {T.~W.}\ \bibnamefont {Neely}},
  \bibinfo {author} {\bibfnamefont {B.~P.}\ \bibnamefont {Anderson}}, \ and\
  \bibinfo {author} {\bibfnamefont {A.~S.}\ \bibnamefont {Bradley}},\ }\href
  {\doibase 10.1103/PhysRevA.88.063620} {\bibfield  {journal} {\bibinfo
  {journal} {Phys. Rev. A}\ }\textbf {\bibinfo {volume} {88}},\ \bibinfo
  {pages} {063620} (\bibinfo {year} {2013})}\BibitemShut {NoStop}%
\bibitem [{\citenamefont {Davis}\ \emph {et~al.}(2013)\citenamefont {Davis},
  \citenamefont {Proukakis}, \citenamefont {Gardiner},\ and\ \citenamefont
  {Szymanska}}]{davis_13}%
  \BibitemOpen
  \bibfield  {author} {\bibinfo {author} {\bibfnamefont {M.}~\bibnamefont
  {Davis}}, \bibinfo {author} {\bibfnamefont {N.}~\bibnamefont {Proukakis}},
  \bibinfo {author} {\bibfnamefont {S.}~\bibnamefont {Gardiner}}, \ and\
  \bibinfo {author} {\bibfnamefont {M.}~\bibnamefont {Szymanska}},\ }\href
  {https://books.google.it/books?id=UjO6CgAAQBAJ} {\emph {\bibinfo {title}
  {Quantum Gases: Finite Temperature and Non-Equilibrium Dynamics}}},\ Cold
  atoms\ (\bibinfo  {publisher} {Imperial College Press},\ \bibinfo {year}
  {2013})\BibitemShut {NoStop}%
\bibitem [{\citenamefont {Berloff}\ \emph {et~al.}(2014)\citenamefont
  {Berloff}, \citenamefont {Brachet},\ and\ \citenamefont
  {Proukakis}}]{berloff_14}%
  \BibitemOpen
  \bibfield  {author} {\bibinfo {author} {\bibfnamefont {N.~G.}\ \bibnamefont
  {Berloff}}, \bibinfo {author} {\bibfnamefont {M.}~\bibnamefont {Brachet}}, \
  and\ \bibinfo {author} {\bibfnamefont {N.~P.}\ \bibnamefont {Proukakis}},\
  }\href {\doibase 10.1073/pnas.1312549111} {\bibfield  {journal} {\bibinfo
  {journal} {PNAS}\ }\textbf {\bibinfo {volume} {111}},\ \bibinfo {pages}
  {4675} (\bibinfo {year} {2014})}\BibitemShut {NoStop}%
\bibitem [{\citenamefont {Brewczyk}\ \emph {et~al.}(2007)\citenamefont
  {Brewczyk}, \citenamefont {Gajda},\ and\ \citenamefont
  {Rza{\.{z}}ewski}}]{brewczyk_2007}%
  \BibitemOpen
  \bibfield  {author} {\bibinfo {author} {\bibfnamefont {M.}~\bibnamefont
  {Brewczyk}}, \bibinfo {author} {\bibfnamefont {M.}~\bibnamefont {Gajda}}, \
  and\ \bibinfo {author} {\bibfnamefont {K.}~\bibnamefont {Rza{\.{z}}ewski}},\
  }\href {\doibase 10.1088/0953-4075/40/2/r01} {\bibfield  {journal} {\bibinfo
  {journal} {J. Phys. B}\ }\textbf {\bibinfo {volume} {40}},\ \bibinfo {pages}
  {R1} (\bibinfo {year} {2007})}\BibitemShut {NoStop}%
\bibitem [{\citenamefont {Gallucci}\ and\ \citenamefont
  {Proukakis}(2016)}]{gallucci_16}%
  \BibitemOpen
  \bibfield  {author} {\bibinfo {author} {\bibfnamefont {D.}~\bibnamefont
  {Gallucci}}\ and\ \bibinfo {author} {\bibfnamefont {N.~P.}\ \bibnamefont
  {Proukakis}},\ }\href {\doibase 10.1088/1367-2630/18/2/025004} {\bibfield
  {journal} {\bibinfo  {journal} {New J. Phys.}\ }\textbf {\bibinfo {volume}
  {18}},\ \bibinfo {pages} {025004} (\bibinfo {year} {2016})}\BibitemShut
  {NoStop}%
\bibitem [{\citenamefont {Kobayashi}\ and\ \citenamefont
  {Cugliandolo}(2016)}]{kobayashi_16}%
  \BibitemOpen
  \bibfield  {author} {\bibinfo {author} {\bibfnamefont {M.}~\bibnamefont
  {Kobayashi}}\ and\ \bibinfo {author} {\bibfnamefont {L.~F.}\ \bibnamefont
  {Cugliandolo}},\ }\href {\doibase 10.1209/0295-5075/115/20007} {\bibfield
  {journal} {\bibinfo  {journal} {{EPL}}\ }\textbf {\bibinfo {volume} {115}},\
  \bibinfo {pages} {20007} (\bibinfo {year} {2016})}\BibitemShut {NoStop}%
\bibitem [{\citenamefont {Ota}\ \emph {et~al.}(2018)\citenamefont {Ota},
  \citenamefont {Larcher}, \citenamefont {Dalfovo}, \citenamefont {Pitaevskii},
  \citenamefont {Proukakis},\ and\ \citenamefont {Stringari}}]{ota_18}%
  \BibitemOpen
  \bibfield  {author} {\bibinfo {author} {\bibfnamefont {M.}~\bibnamefont
  {Ota}}, \bibinfo {author} {\bibfnamefont {F.}~\bibnamefont {Larcher}},
  \bibinfo {author} {\bibfnamefont {F.}~\bibnamefont {Dalfovo}}, \bibinfo
  {author} {\bibfnamefont {L.}~\bibnamefont {Pitaevskii}}, \bibinfo {author}
  {\bibfnamefont {N.~P.}\ \bibnamefont {Proukakis}}, \ and\ \bibinfo {author}
  {\bibfnamefont {S.}~\bibnamefont {Stringari}},\ }\href {\doibase
  10.1103/PhysRevLett.121.145302} {\bibfield  {journal} {\bibinfo  {journal}
  {Phys. Rev. Lett.}\ }\textbf {\bibinfo {volume} {121}},\ \bibinfo {pages}
  {145302} (\bibinfo {year} {2018})}\BibitemShut {NoStop}%
\bibitem [{\citenamefont {Bradley}\ and\ \citenamefont
  {Blakie}(2014)}]{bradley_14}%
  \BibitemOpen
  \bibfield  {author} {\bibinfo {author} {\bibfnamefont {A.~S.}\ \bibnamefont
  {Bradley}}\ and\ \bibinfo {author} {\bibfnamefont {P.~B.}\ \bibnamefont
  {Blakie}},\ }\href {\doibase 10.1103/PhysRevA.90.023631} {\bibfield
  {journal} {\bibinfo  {journal} {Phys. Rev. A}\ }\textbf {\bibinfo {volume}
  {90}},\ \bibinfo {pages} {023631} (\bibinfo {year} {2014})}\BibitemShut
  {NoStop}%
\bibitem [{\citenamefont {Rooney}\ \emph {et~al.}(2010)\citenamefont {Rooney},
  \citenamefont {Bradley},\ and\ \citenamefont {Blakie}}]{rooney_10}%
  \BibitemOpen
  \bibfield  {author} {\bibinfo {author} {\bibfnamefont {S.~J.}\ \bibnamefont
  {Rooney}}, \bibinfo {author} {\bibfnamefont {A.~S.}\ \bibnamefont {Bradley}},
  \ and\ \bibinfo {author} {\bibfnamefont {P.~B.}\ \bibnamefont {Blakie}},\
  }\href {\doibase 10.1103/PhysRevA.81.023630} {\bibfield  {journal} {\bibinfo
  {journal} {Phys. Rev. A}\ }\textbf {\bibinfo {volume} {81}},\ \bibinfo
  {pages} {023630} (\bibinfo {year} {2010})}\BibitemShut {NoStop}%
\bibitem [{\citenamefont {Comaron}\ \emph {et~al.}(2019)\citenamefont
  {Comaron}, \citenamefont {Larcher}, \citenamefont {Dalfovo},\ and\
  \citenamefont {Proukakis}}]{comaron_19}%
  \BibitemOpen
  \bibfield  {author} {\bibinfo {author} {\bibfnamefont {P.}~\bibnamefont
  {Comaron}}, \bibinfo {author} {\bibfnamefont {F.}~\bibnamefont {Larcher}},
  \bibinfo {author} {\bibfnamefont {F.}~\bibnamefont {Dalfovo}}, \ and\
  \bibinfo {author} {\bibfnamefont {N.~P.}\ \bibnamefont {Proukakis}},\ }\href
  {\doibase 10.1103/PhysRevA.100.033618} {\bibfield  {journal} {\bibinfo
  {journal} {Phys. Rev. A}\ }\textbf {\bibinfo {volume} {100}},\ \bibinfo
  {pages} {033618} (\bibinfo {year} {2019})}\BibitemShut {NoStop}%
\bibitem [{\citenamefont {Larcher}(2018)}]{fabrizio_18}%
  \BibitemOpen
  \bibfield  {author} {\bibinfo {author} {\bibfnamefont {F.}~\bibnamefont
  {Larcher}},\ }\emph {\bibinfo {title} {Dynamical excitations in
  low-dimensional condensates: sound, vortices and quenched dynamics}},\ \href
  {http://eprints-phd.biblio.unitn.it/2902/} {Ph.D. thesis},\ \bibinfo
  {school} {University of Trento and Newcastle University.} (\bibinfo {year}
  {2018})\BibitemShut {NoStop}%
\bibitem [{\citenamefont {Liu}\ \emph {et~al.}(2020)\citenamefont {Liu},
  \citenamefont {Dziarmaga}, \citenamefont {Gou}, \citenamefont {Dalfovo},\
  and\ \citenamefont {Proukakis}}]{liu_20}%
  \BibitemOpen
  \bibfield  {author} {\bibinfo {author} {\bibfnamefont {I.-K.}\ \bibnamefont
  {Liu}}, \bibinfo {author} {\bibfnamefont {J.}~\bibnamefont {Dziarmaga}},
  \bibinfo {author} {\bibfnamefont {S.-C.}\ \bibnamefont {Gou}}, \bibinfo
  {author} {\bibfnamefont {F.}~\bibnamefont {Dalfovo}}, \ and\ \bibinfo
  {author} {\bibfnamefont {N.~P.}\ \bibnamefont {Proukakis}},\ }\href {\doibase
  10.1103/PhysRevResearch.2.033183} {\bibfield  {journal} {\bibinfo  {journal}
  {Phys. Rev. Research}\ }\textbf {\bibinfo {volume} {2}},\ \bibinfo {pages}
  {033183} (\bibinfo {year} {2020})}\BibitemShut {NoStop}%
\bibitem [{\citenamefont {Liu}\ \emph {et~al.}(2018)\citenamefont {Liu},
  \citenamefont {Donadello}, \citenamefont {Lamporesi}, \citenamefont
  {Ferrari}, \citenamefont {Gou}, \citenamefont {Dalfovo},\ and\ \citenamefont
  {Proukakis}}]{liu2018}%
  \BibitemOpen
  \bibfield  {author} {\bibinfo {author} {\bibfnamefont {I.-K.}\ \bibnamefont
  {Liu}}, \bibinfo {author} {\bibfnamefont {S.}~\bibnamefont {Donadello}},
  \bibinfo {author} {\bibfnamefont {G.}~\bibnamefont {Lamporesi}}, \bibinfo
  {author} {\bibfnamefont {G.}~\bibnamefont {Ferrari}}, \bibinfo {author}
  {\bibfnamefont {S.-C.}\ \bibnamefont {Gou}}, \bibinfo {author} {\bibfnamefont
  {F.}~\bibnamefont {Dalfovo}}, \ and\ \bibinfo {author} {\bibfnamefont
  {N.}~\bibnamefont {Proukakis}},\ }\href
  {https://www.nature.com/articles/s42005-018-0023-6} {\bibfield  {journal}
  {\bibinfo  {journal} {Communications Physics}\ }\textbf {\bibinfo {volume}
  {1}},\ \bibinfo {pages} {24} (\bibinfo {year} {2018})}\BibitemShut {NoStop}%
\bibitem [{\citenamefont {Sakmann}\ and\ \citenamefont
  {Kasevich}(2016)}]{sakmann_16}%
  \BibitemOpen
  \bibfield  {author} {\bibinfo {author} {\bibfnamefont {K.}~\bibnamefont
  {Sakmann}}\ and\ \bibinfo {author} {\bibfnamefont {M.}~\bibnamefont
  {Kasevich}},\ }\href {\doibase 10.1038/nphys3631} {\bibfield  {journal}
  {\bibinfo  {journal} {Nature Physics}\ }\textbf {\bibinfo {volume} {12}},\
  \bibinfo {pages} {451} (\bibinfo {year} {2016})}\BibitemShut {NoStop}%
\bibitem [{\citenamefont {Sakmann}\ and\ \citenamefont
  {Kasevich}(2017)}]{sakmann_17}%
  \BibitemOpen
  \bibfield  {author} {\bibinfo {author} {\bibfnamefont {K.}~\bibnamefont
  {Sakmann}}\ and\ \bibinfo {author} {\bibfnamefont {M.}~\bibnamefont
  {Kasevich}},\ }\href {\doibase 10.48550/ARXIV.1702.01211} {\bibfield
  {journal} {\bibinfo  {journal} {arXiv:1702.01211}\ } (\bibinfo {year}
  {2017}),\ 10.48550/ARXIV.1702.01211}\BibitemShut {NoStop}%
\bibitem [{\citenamefont {Olsen}\ \emph {et~al.}(2017)\citenamefont {Olsen},
  \citenamefont {Corney}, \citenamefont {Lewis-Swan},\ and\ \citenamefont
  {Bradley}}]{olsen-17a}%
  \BibitemOpen
  \bibfield  {author} {\bibinfo {author} {\bibfnamefont {M.~K.}\ \bibnamefont
  {Olsen}}, \bibinfo {author} {\bibfnamefont {J.~F.}\ \bibnamefont {Corney}},
  \bibinfo {author} {\bibfnamefont {R.~J.}\ \bibnamefont {Lewis-Swan}}, \ and\
  \bibinfo {author} {\bibfnamefont {A.~S.}\ \bibnamefont {Bradley}},\ }\href
  {\doibase 10.48550/ARXIV.1702.00282} {\bibfield  {journal} {\bibinfo
  {journal} {arXiv:1702.00282}\ } (\bibinfo {year} {2017}),\
  10.48550/ARXIV.1702.00282}\BibitemShut {NoStop}%
\bibitem [{\citenamefont {Ville}\ \emph {et~al.}(2018)\citenamefont {Ville},
  \citenamefont {Saint-Jalm}, \citenamefont {Le~Cerf}, \citenamefont
  {Aidelsburger}, \citenamefont {Nascimb\`ene}, \citenamefont {Dalibard},\ and\
  \citenamefont {Beugnon}}]{ville_18}%
  \BibitemOpen
  \bibfield  {author} {\bibinfo {author} {\bibfnamefont {J.~L.}\ \bibnamefont
  {Ville}}, \bibinfo {author} {\bibfnamefont {R.}~\bibnamefont {Saint-Jalm}},
  \bibinfo {author} {\bibfnamefont {E.}~\bibnamefont {Le~Cerf}}, \bibinfo
  {author} {\bibfnamefont {M.}~\bibnamefont {Aidelsburger}}, \bibinfo {author}
  {\bibfnamefont {S.}~\bibnamefont {Nascimb\`ene}}, \bibinfo {author}
  {\bibfnamefont {J.}~\bibnamefont {Dalibard}}, \ and\ \bibinfo {author}
  {\bibfnamefont {J.}~\bibnamefont {Beugnon}},\ }\href {\doibase
  10.1103/PhysRevLett.121.145301} {\bibfield  {journal} {\bibinfo  {journal}
  {Phys. Rev. Lett.}\ }\textbf {\bibinfo {volume} {121}},\ \bibinfo {pages}
  {145301} (\bibinfo {year} {2018})}\BibitemShut {NoStop}%
\bibitem [{\citenamefont {Cominotti}\ \emph
  {et~al.}(2022{\natexlab{b}})\citenamefont {Cominotti}, \citenamefont {Berti},
  \citenamefont {Farolfi}, \citenamefont {Zenesini}, \citenamefont {Lamporesi},
  \citenamefont {Carusotto}, \citenamefont {Recati},\ and\ \citenamefont
  {Ferrari}}]{cominotti_21}%
  \BibitemOpen
  \bibfield  {author} {\bibinfo {author} {\bibfnamefont {R.}~\bibnamefont
  {Cominotti}}, \bibinfo {author} {\bibfnamefont {A.}~\bibnamefont {Berti}},
  \bibinfo {author} {\bibfnamefont {A.}~\bibnamefont {Farolfi}}, \bibinfo
  {author} {\bibfnamefont {A.}~\bibnamefont {Zenesini}}, \bibinfo {author}
  {\bibfnamefont {G.}~\bibnamefont {Lamporesi}}, \bibinfo {author}
  {\bibfnamefont {I.}~\bibnamefont {Carusotto}}, \bibinfo {author}
  {\bibfnamefont {A.}~\bibnamefont {Recati}}, \ and\ \bibinfo {author}
  {\bibfnamefont {G.}~\bibnamefont {Ferrari}},\ }\href {\doibase
  10.1103/PhysRevLett.128.210401} {\bibfield  {journal} {\bibinfo  {journal}
  {Phys. Rev. Lett.}\ }\textbf {\bibinfo {volume} {128}},\ \bibinfo {pages}
  {210401} (\bibinfo {year} {2022}{\natexlab{b}})}\BibitemShut {NoStop}%
\bibitem [{\citenamefont {Prokof'ev}\ \emph {et~al.}(2001)\citenamefont
  {Prokof'ev}, \citenamefont {Ruebenacker},\ and\ \citenamefont
  {Svistunov}}]{prokofev_01}%
  \BibitemOpen
  \bibfield  {author} {\bibinfo {author} {\bibfnamefont {N.}~\bibnamefont
  {Prokof'ev}}, \bibinfo {author} {\bibfnamefont {O.}~\bibnamefont
  {Ruebenacker}}, \ and\ \bibinfo {author} {\bibfnamefont {B.}~\bibnamefont
  {Svistunov}},\ }\href {\doibase 10.1103/PhysRevLett.87.270402} {\bibfield
  {journal} {\bibinfo  {journal} {Phys. Rev. Lett.}\ }\textbf {\bibinfo
  {volume} {87}},\ \bibinfo {pages} {270402} (\bibinfo {year}
  {2001})}\BibitemShut {NoStop}%
\bibitem [{\citenamefont {Kristensen}\ \emph {et~al.}(2019)\citenamefont
  {Kristensen}, \citenamefont {Christensen}, \citenamefont {Gajdacz},
  \citenamefont {Iglicki}, \citenamefont {Pawlowski}, \citenamefont {Klempt},
  \citenamefont {Sherson}, \citenamefont {Rzazewski}, \citenamefont
  {Hilliard},\ and\ \citenamefont {Arlt}}]{kristensen_19}%
  \BibitemOpen
  \bibfield  {author} {\bibinfo {author} {\bibfnamefont {M.~A.}\ \bibnamefont
  {Kristensen}}, \bibinfo {author} {\bibfnamefont {M.~B.}\ \bibnamefont
  {Christensen}}, \bibinfo {author} {\bibfnamefont {M.}~\bibnamefont
  {Gajdacz}}, \bibinfo {author} {\bibfnamefont {M.}~\bibnamefont {Iglicki}},
  \bibinfo {author} {\bibfnamefont {K.}~\bibnamefont {Pawlowski}}, \bibinfo
  {author} {\bibfnamefont {C.}~\bibnamefont {Klempt}}, \bibinfo {author}
  {\bibfnamefont {J.~F.}\ \bibnamefont {Sherson}}, \bibinfo {author}
  {\bibfnamefont {K.}~\bibnamefont {Rzazewski}}, \bibinfo {author}
  {\bibfnamefont {A.~J.}\ \bibnamefont {Hilliard}}, \ and\ \bibinfo {author}
  {\bibfnamefont {J.~J.}\ \bibnamefont {Arlt}},\ }\href {\doibase
  10.1103/PhysRevLett.122.163601} {\bibfield  {journal} {\bibinfo  {journal}
  {Phys. Rev. Lett.}\ }\textbf {\bibinfo {volume} {122}},\ \bibinfo {pages}
  {163601} (\bibinfo {year} {2019})}\BibitemShut {NoStop}%
\bibitem [{\citenamefont {Continentino}(2017)}]{Continentino}%
  \BibitemOpen
  \bibfield  {author} {\bibinfo {author} {\bibfnamefont {M.}~\bibnamefont
  {Continentino}},\ }\href@noop {} {\emph {\bibinfo {title} {Quantum Scaling in
  Many-Body Systems}}}\ (\bibinfo  {publisher} {Cambridge University Press},\
  \bibinfo {year} {2017})\BibitemShut {NoStop}%
\bibitem [{\citenamefont {Hohenberg}\ and\ \citenamefont
  {Halperin}(1977)}]{hohenberg_77}%
  \BibitemOpen
  \bibfield  {author} {\bibinfo {author} {\bibfnamefont {P.~C.}\ \bibnamefont
  {Hohenberg}}\ and\ \bibinfo {author} {\bibfnamefont {B.~I.}\ \bibnamefont
  {Halperin}},\ }\href {\doibase 10.1103/RevModPhys.49.435} {\bibfield
  {journal} {\bibinfo  {journal} {Rev. Mod. Phys.}\ }\textbf {\bibinfo {volume}
  {49}},\ \bibinfo {pages} {435} (\bibinfo {year} {1977})}\BibitemShut
  {NoStop}%
\bibitem [{not()}]{notefactor2}%
  \BibitemOpen
  \href@noop {} {}\bibinfo {note} {The notation in
  ref.~\protect\cite{recati_21} is slightly different; in particular, it
  differs by a factor $2$ in the definition of $\Omega$.}\BibitemShut {Stop}%
\end{thebibliography}%
\bibliographystyle{apsrev4-1}

\end{document}